# Mass Distribution, Rotation Curves and Gravity Theories

**Rahul Datta**
Space Applications Centre, Indian Space Research Organization,
Ahmedabad-380015, India
**Dilip G. Banhatti**
School of Physics, Madurai-Kamaraj University, Madurai 625021, India
[Email: dilip.g.banhatti@gmail.com]

## ABSTRACT

**{Comparison of mass density profiles of galaxies of varying sizes from some gravity theories based on observed galaxy rotation curves and assessing the need for dark matter.}**

We present an analysis of five galaxies of varying galactic radii: NGC6822 (4.8 kpc), Large Magellanic Cloud (9 kpc), Milky Way (17 kpc), NGC3198 (30 kpc) and UGC9133 (102.5 kpc). The masses and mass density profiles of these galaxies have been computed using the scientific computing s/w package MATLAB taking the already available velocity profiles of the galaxies as the input, and without considering any dark matter contribution. We have plotted these profiles after computing them according to three different theories of gravity (and dynamics) : Newtonian (black line), Modified Newtonian Dynamics (MoND) (green line) and Vacuum Modified Gravity (red line). We also consider how the profile due to Newtonian theory would modify if we take into account a small negative value of the cosmological constant (-5 x $10^{-56}$ $cm^{-2}$ from theory) (blue line). Comparing these masses and mass density profiles, we try to form an idea regarding what could be a realistic theory of gravity and whether we need dark matter to explain the results.

**Keywords** : disk galaxy rotation curves, galaxy mass, mass density profile, dark matter, Newtonian theory, MoND, Vacuum Modified Gravity, negative cosmological constant

## MOTIVATION

“Missing mass” was first inferred via dynamics for clusters of mostly elliptical galaxies using virial theorem. It was soon found to be dynamically present, although it was the light that was missing, same as in clusters, on even single galaxy scale, especially for spirals, most naturally taken to be balanced between rotation and gravity. Assigning mass-to-light ratios on various scales via putative astrophysical light emission processes on those scales, *dark matter*, which gravitated (and hence participated in dynamics), but emitted little or no light, became firmly established in astrophysics and cosmology.

In this paper, we consider the single galaxy scale, and use five galaxies, either disk spirals or dynamically similar galaxies, to examine the connection between mass distribution and rotation curve in a few representative models, without assuming any dark matter, on scales from 5 kpc (size of the smallest of the five galaxies) to 100 kpc (size of the largest of the 5).

## Calculation / Computation Scheme

We first define the geometry and the notation for our calculation (computed using MATLAB s/w package). The galaxy disk is assumed axisymmetric.

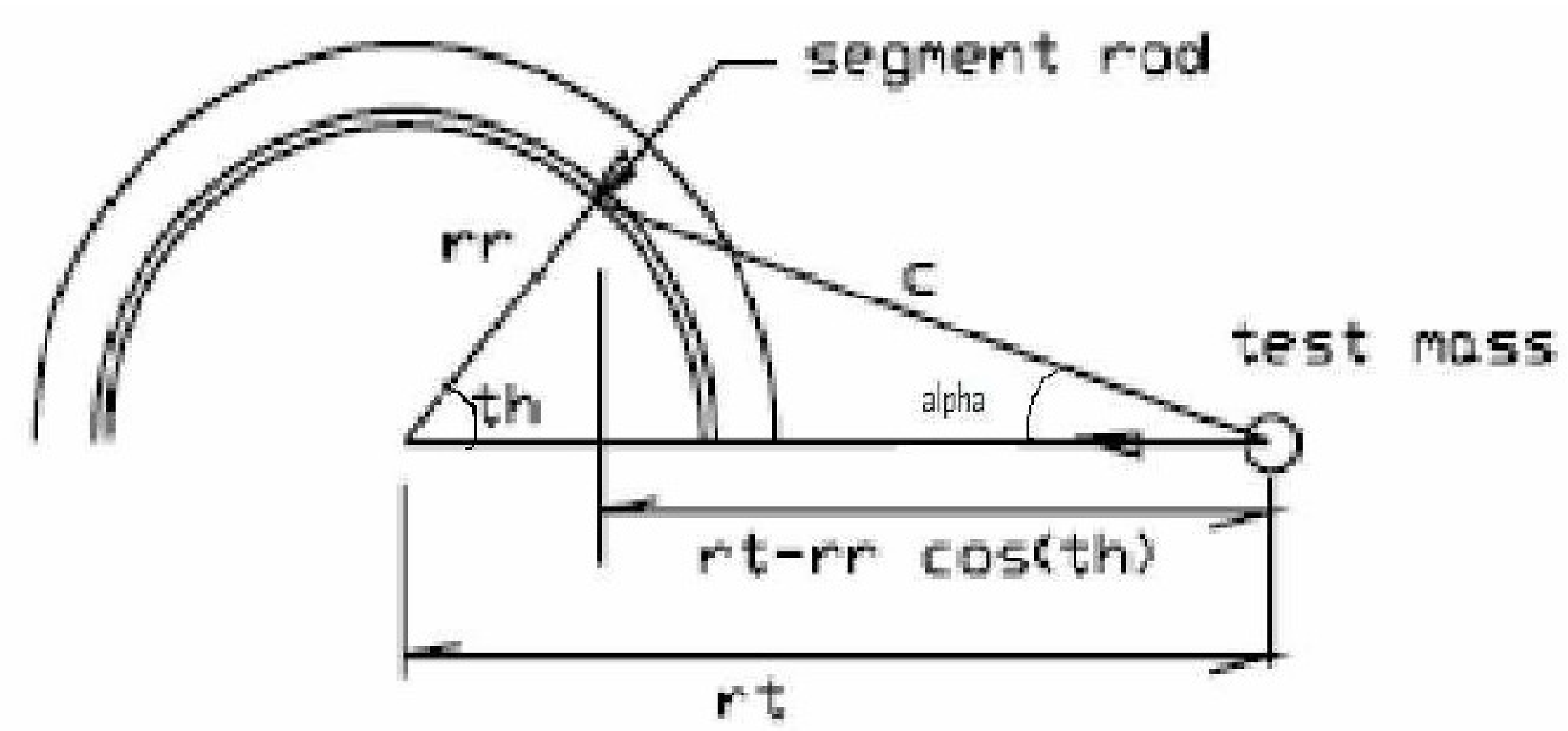

FIG.1: Top view of a galaxy. The semicircular arcs represent rings constituting the disk. The other half forming the full circular arc is implied.

Definitions & notation (referring to FIGs. 1 & 2):
pc = parsec, 3.08568E13 kms, 3.26151 light years
pi = 3.14159265
dr = radial thickness of a ring, pc
dth = (pi /180) radians
th = angle at the galaxy center, between galaxy radial line to test mass and radial line to ring segment, radians
alpha = angle subtended at the test mass by the center of the fundamental segment mass and the center of the galaxy, radians
beta = angle subtended at the test mass by the center of the fundamental segment mass and the elemental thickness dz of the fundamental segment mass, radians
pytks = multiplier to change v (pc/yr) to v (km/sec) = 9.7778E5 km/pc*yr/sec
msun = mass of our sun, 1.989E33 gms
htype indicates type of the galaxy

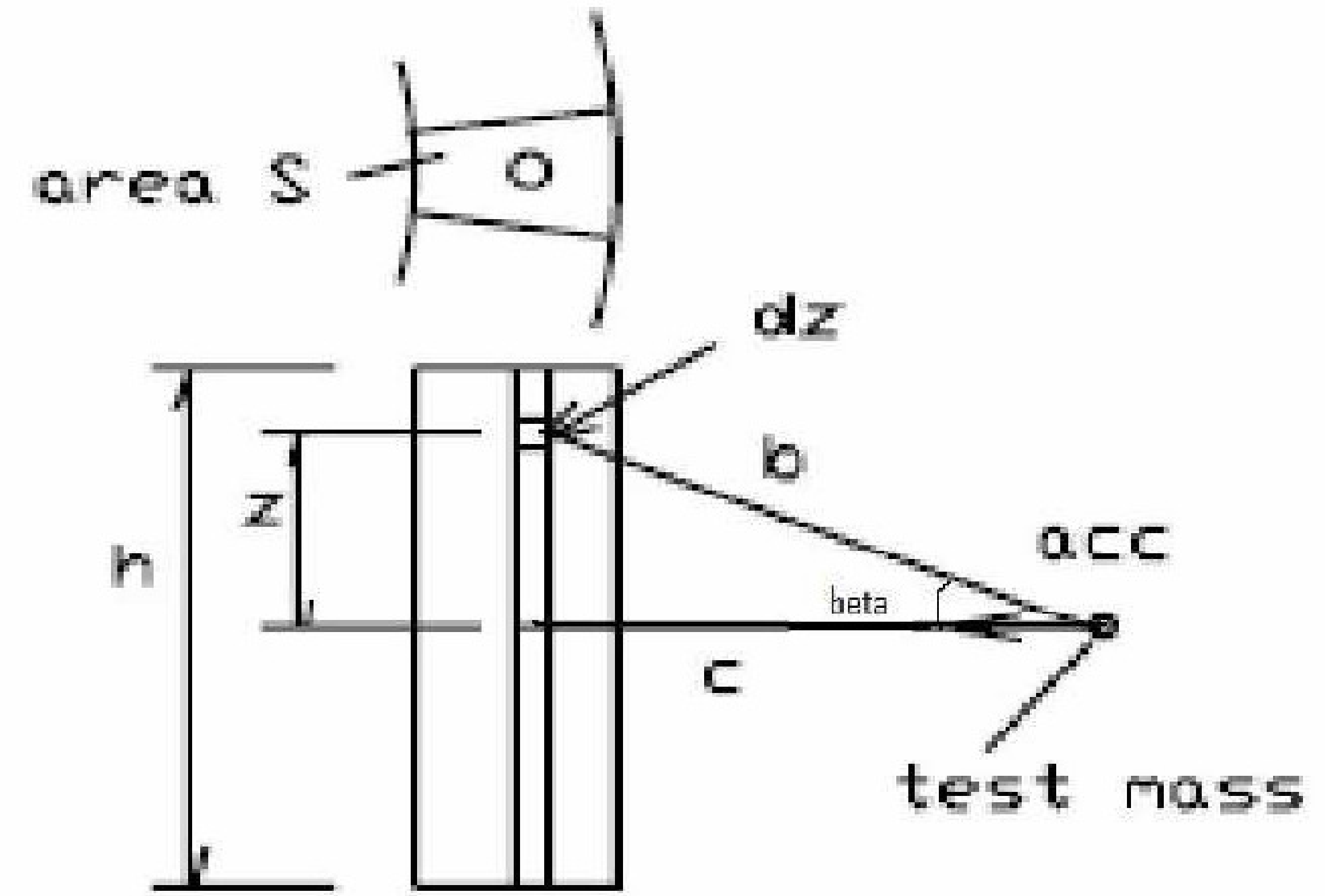

FIG.2: Edge-on view of a fundamental segment of a ring in an axisymmetric disk galaxy. Cf FIG.1.

r(i) = radius to centerline circle of ith ring = rm - dr/2 pc
rr(i) = radius to rod used to represent fundamental segment mass for the ith ring = rm – (dr / 2) * (rm - 0.575 dr) / (rm - dr / 2) pc
Nr = number of rings
G = gravitational constant, 4.498E(-15) pc^3/(msuns*yr^2)
rm = radius to outer edge of a ring, pc
rmax = galaxy maximum/rim radius, pc
rt = radius to test mass from galaxy center, pcs
the test mass is located at the outer edge of each ring
h(i) = galaxy thickness at radius r corresponding to ith ring, pc
mr(i) = mass contained only inside the ith ring, msuns
Mr(i) = galaxy mass contained within the outer edge of the ith ring, i.e., sum of masses inside each ring starting from the galaxy center upto the ith ring, msuns
M = galaxy total mass, msuns
dv(i) = volume of a fundamental segment of the ith ring, pc^3
        = h(i)*r(i)*dth*dr
rv(i) = volume of the ith ring, pc^3
        = 360*dv(i)

tvol = galaxy total volume, pc^3
rho(i) = density of the ith ring, msuns/pc^3
rhoav = average density of galaxy = M / tvol, msuns/pc^3
SMD(i) = surface mass density of ith ring, msuns / pc^2

SMDav = average surface mass density of galaxy
= M / ( $\pi$ * rmax2 ) msuns / pc^2
rSMD(i) = rd(i) * SMD(i)
v(i) = computed orbital speed at test mass radius rt at the outer edge of ith ring, kms/sec
vkrim = sqrt(GM / rmax) , the Kepler speed at the rim, kms/sec
vm = measured speed, km/sec
vmax = maximum measured speed, km/sec
mrnor(i) = mr(i) / M
Mrnor(i) = Mr(i) / M
hd(i), rd(i) = (h(i), r(i)) / rmax
rhonor(i) = rho(i) / rhoav
SMDnor(i) = SMD(i) / SMDav
rSMDnor(i) = rd(i) * SMDnor(i)
vnor(i) = v(i) / vkrim

FORWARD PROBLEM: Computation of the rotation profile from a given mass distribution.

Referring to FIG.2, acceleration due to one fundamental segment mass of a ring: From Newton's law, centripetal acceleration = $GM/r^2$
Here, for a fundamental segment, rho is a constant.
Hence, $M/r^2$ is equivalent to rho $\int (Sdz/b^2) \cos(beta)$, taking the component of the acceleration due to the mass of the elemental thickness, dz of the fundamental segment of a ring along the direction of the plane of the galactic disc, where S = cross-sectional area of the fundamental segment cut along the plane of the galaxy disc. Therefore,
Acc = $G (rho) \int (S\, dz/b^2)\cos(beta)$
$= G (rho) \int (S\, dz/b^2)(c/b)$
But, $z = c \tan(beta)$ and $b = c \sec(beta)$ or, $dz = c \sec^2(beta)\, d(beta)$. Therefore, Acc
$= G (rho) S \int (c \sec^2(beta)\, d(beta)/b^2)(c/b)$
$= G (rho) S \int (c^2 \sec^2(beta)/b^3) d(beta)$
$= G (rho) S \int (c^2 \sec^2(beta)/c^3 \sec^3(beta)) d(beta)$
$= G (rho)(S/c) \int d(beta)/\sec(beta)$

$$= G\,(rho)(S/c) \int \cos(beta)d(beta)$$

$$= G\,(rho)(S/c)\,[\sin(beta)]_{z=-h/2}^{z=+h/2}$$

$$= G\,(rho)(S/c)\,[z/(c^2+z^2)^{0.5}]_{z=-h/2}^{z=+h/2}$$

$$= G\,(rho)(S/c)\,[(h/2)/(c^2+(h/2)^2)^{0.5} + (h/2)/(c^2+(h/2)^2)^{0.5}]$$

$$= G\,(rho)(S/c)\,[h/(c^2+(h/2)^2)^{0.5}]$$

Referring to FIG.1, each ring consists of 360 such fundamental segments of mass and angle th varies from 0 to $\pi$ radians or 0 to 180 degrees, both in the clockwise and the anticlockwise direction from the line joining the galaxy center to the test mass. The centers of the fundamental segments correspond to 0.5 degrees, 1.5 degrees, 2.5 degrees and so on until 179.5 degrees both in the clockwise and anticlockwise direction. Therefore the centripetal acceleration of the test mass due to each ring:

$$v^2/rt = 2\,(pytks)^2 \int_{th=0}^{th=\pi} G\,(rho)(S/c)\,[h/(c^2+(h/2)^2)^{0.5}]\cos(alpha)\,dth$$

the cos(alpha) term being multiplied for taking the component of the acceleration of the test mass due to each fundamental segment of the ring along the radial direction towards the galactic center and the term 2 being multiplied to account for both the clockwise and anticlockwise directions.

cos(alpha) = (rt – rr cos(th))/c.

Hence, the squared orbital velocity due to each ring:

$$v^2 = 2\,rt\,(pytks)^2 \int_{th=0}^{th=\pi} G\,(rho)\,(S/c^2)\,[h/(c^2+(h/2)^2)^{0.5}]\,(rt - rr\cos(th))\,dth$$

Discretizing the above formula by converting the integral to summation for the purpose of computation, we have:

$$v^2 = 2\,rt\,(pytks)^2 \sum_{i=0}^{i=180} G\,(rho)\,(S/c^2)\,[h/(c^2+(h/2)^2)^{0.5}]\,(rt - rr\cos((i-0.5)dth))$$

where dth = (pi/180) radians and

rr = (j dr) - (dr/2) {(j dr)-(0.575 dr)/(j dr) - (0.5 dr)}

dr = radial thickness of each ring

S = r(j) dth dr

Summing up over all the rings, we have

$$v^2 = rt\,(pytks)^2 \sum_{j=0}^{j=Nr} 2 \sum_{i=0}^{i=180} G(rho)r(j)dth\ dr/c^2)[h(j)/(c^2+(h(j)/2)^2)^{0.5}][rt - rr\cos((i-0.5)dth)]$$

REVERSE PROBLEM: Finding the mass distribution for a given rotation profile

The reverse problem is done by repeated applications of the forward problem using the equivalent thickness distribution, an arbitrary density distribution is input, and the speed profile computed using the above formula. The speed errors (measured minus computed) at each of the ring outer radii are then used to correct the densities of the rings and the process repeated until all speed errors are small enough to be neglected.

errv = (vm-v) / vmax ; f = 0.75*errv
if all errv < 1E-6 then quit
else
rho(i) = (1+f) rho(i-1) for each cycle i
where vm = measured speed, and vmax = maximum computed speed

To compute the mass and mass density profiles after taking into account the contribution of a cosmological constant, $\lambda$ taken from theory to be equal to -5 x $10^{-56}cm^{-2}$ in magnitude can be found out by using a new velocity profile as input, which is obtained after subtracting a quantity equal to c*r(i)*sqrt($\lambda$/3) from the observed velocity profile. Note that we have considered a negative cosmological constant scenario.

To compute the mass and mass density profiles as per the Modified Newtonian Dynamics (MoND), we use the following equation for the FORWARD Algorithm and the REVERSE Algorithm, which incorporates the results of the Forward Algorithm, remains otherwise the same as earlier.

$$v_{MoND}^2 = v^2\,(1 + \mathrm{sqrt}[(M0/Mr1(j))\ (1 - \exp(-rt/r0)\ (1+rt/r0))])$$

$$v_{MoND}^2 = rt\,(pytks)^2 \sum_{j=0}^{j=Nr} 2 \sum_{i=0}^{i=180} G\,(rho)\ r(j)\ dth\ dr/c^2)[h(j)/(c^2+(h(j)/2)^2)^{0.5}]*$$
$$*[rt - rr\cos((i - 0.5)dth)]\ (1 + \mathrm{sqrt}[(M0/Mr1(j))\ (1 - \exp(-rt/r0)\ (1+rt/r0))])$$

where,

M0 = 9.6e+11 msuns for rmax>12000pc
 = 2.4e+11 msuns for rmax=<12000pc
r0 = 13920 pc for rmax>12000pc
 = 6960 pc for rmax=<12000pc
Mr1(j) = the quantity of mass contained within the radius at which the velocity is being computed.

To compute the mass and mass density profiles as per Vacuum Modified Gravity (VMG), we use the following equation for the FORWARD Algorithm and the REVERSE Algorithm, which incorporates the results of the Forward Algorithm, remains otherwise the same as earlier.

$$v^2 = rt\,(pytks)^2\{ \sum_{j=0}^{j=Nr} 2 \sum_{i=0}^{i=180} (G\ rho\ r(j)\ dth\ dr/c^2)\ [h(j)/(c^2+(h(j)/2)^2)^{0.5}]*$$

$$*[rt - rr\cos((i - 0.5)dth)](\cos(rt/rc1(x))+\sin(rt/rc1(x)))$$

$$- \sum_{j=0}^{j=Nr} 2 \sum_{i=0}^{i=180} (G\ rho\ r(j)\ dth\ dr)\ [\mathrm{atan}(4\,c\,h(j)/(4\,c^2 - h(j)^2))/c]*$$

$$*[rt - rr\cos((i-0.5)dth)]\ (\cos(rt/rc1(j))-\sin(rt/rc1(j)))/rc1(j)\ \}$$

where
$rhov1(j)=(1/3)*(vel)^4/4/3.14159265/Mr1(j)/G/beta$
$rc1(j)=(vel)^2*sqrt(1/(3*4*3.14159265*rhov1(j)*G))$
Mr1(j) = the quantity of mass contained within the radius at which the velocity is being computed.
$beta=1.22*1.22*(1.989E+30)/(3.08568E+16)^2$ pc/sqrt(msuns)
vel=3/9.7778 pc/year

## Results

Nine figures with self-explanatory captions are presented for each galaxy, followed by a table summarizing integral features of the galaxy. Thus there are 45 figures and 5 tables in all. The galaxies, their sizes & pages with their results are as follows:

NGC 6822……………………………………..4.8 kpc……….pp 8 to 17
Large Magellanic Cloud (LMC)………………9 kpc…………pp 28 to 37
Milky Way……………………………………17 kpc………...pp 38 to 47
NGC 3198……………………………………..30 kpc………...pp 48 to 57
UGC 9133……………………………………..102.5 kpc……..pp 18 to 27

**NGC 6822 (4.8 kpc)**

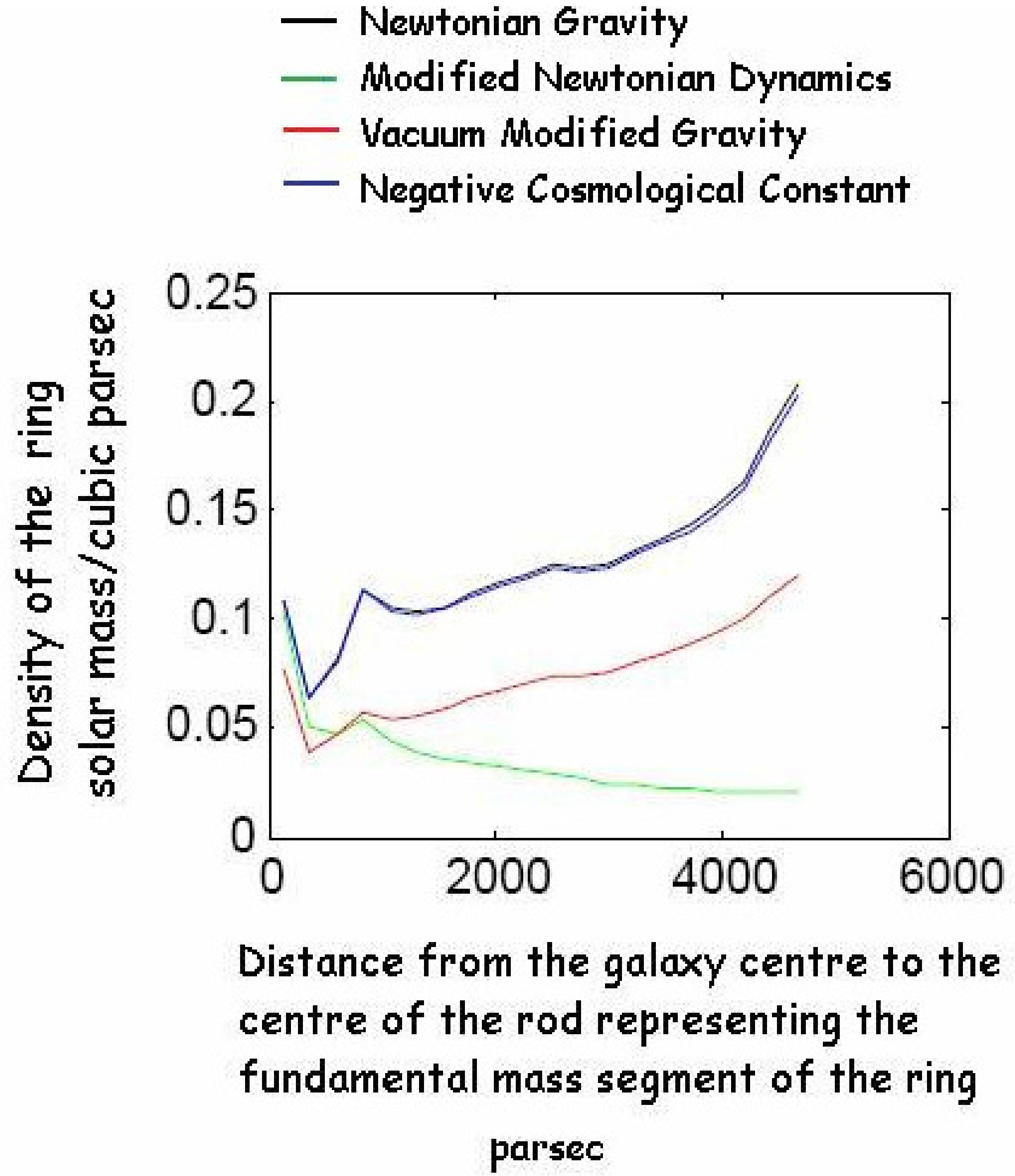

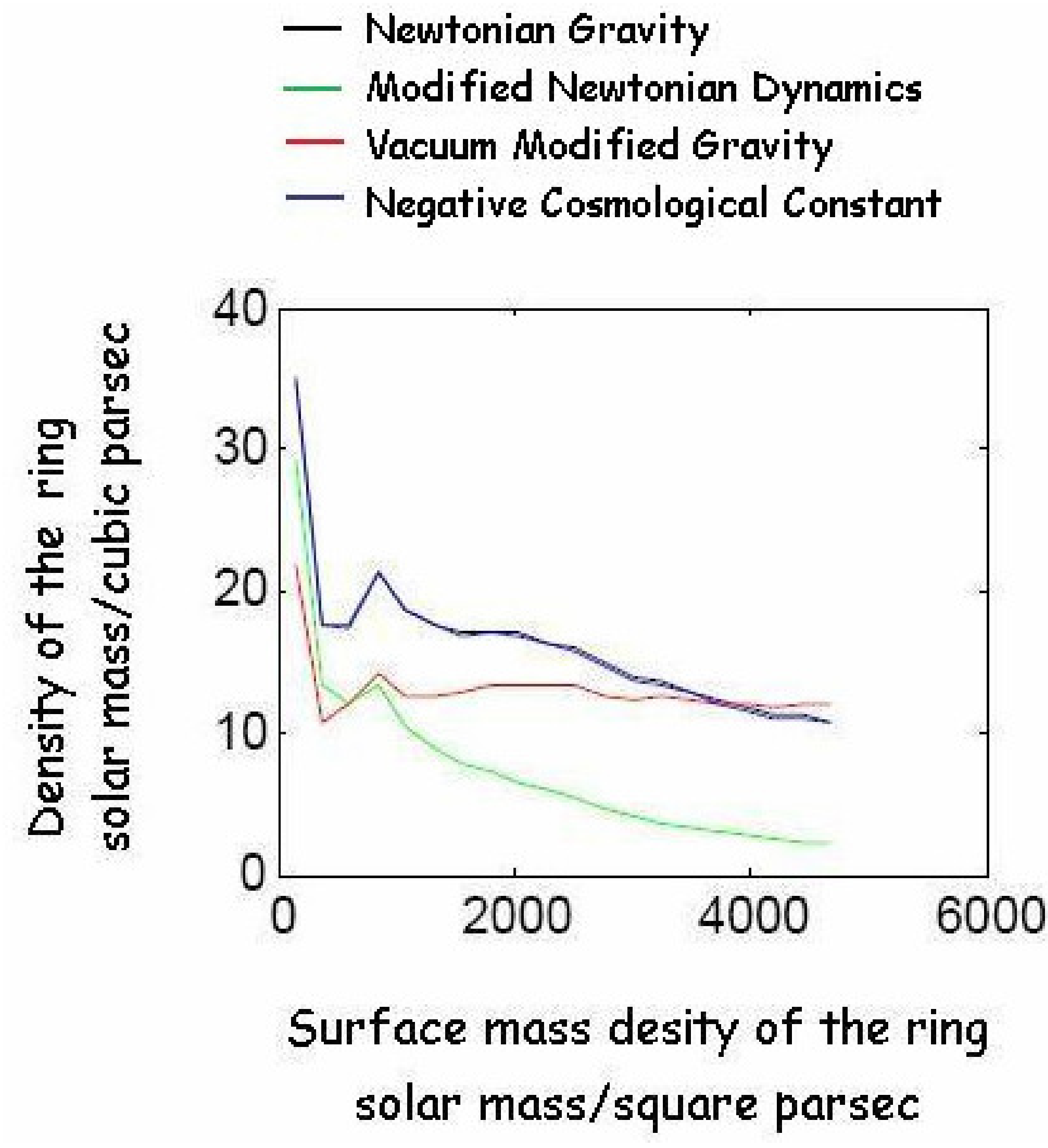

Newtonian Gravity
Modified Newtonian Dynamics
Vacuum Modified Gravity
Negative Cosmological Constant
40
30
20
10
0
0
2000
4000
6000
Density of the ring
solar mass/cubic parsec
Surface mass desity of the ring
solar mass/square parsec

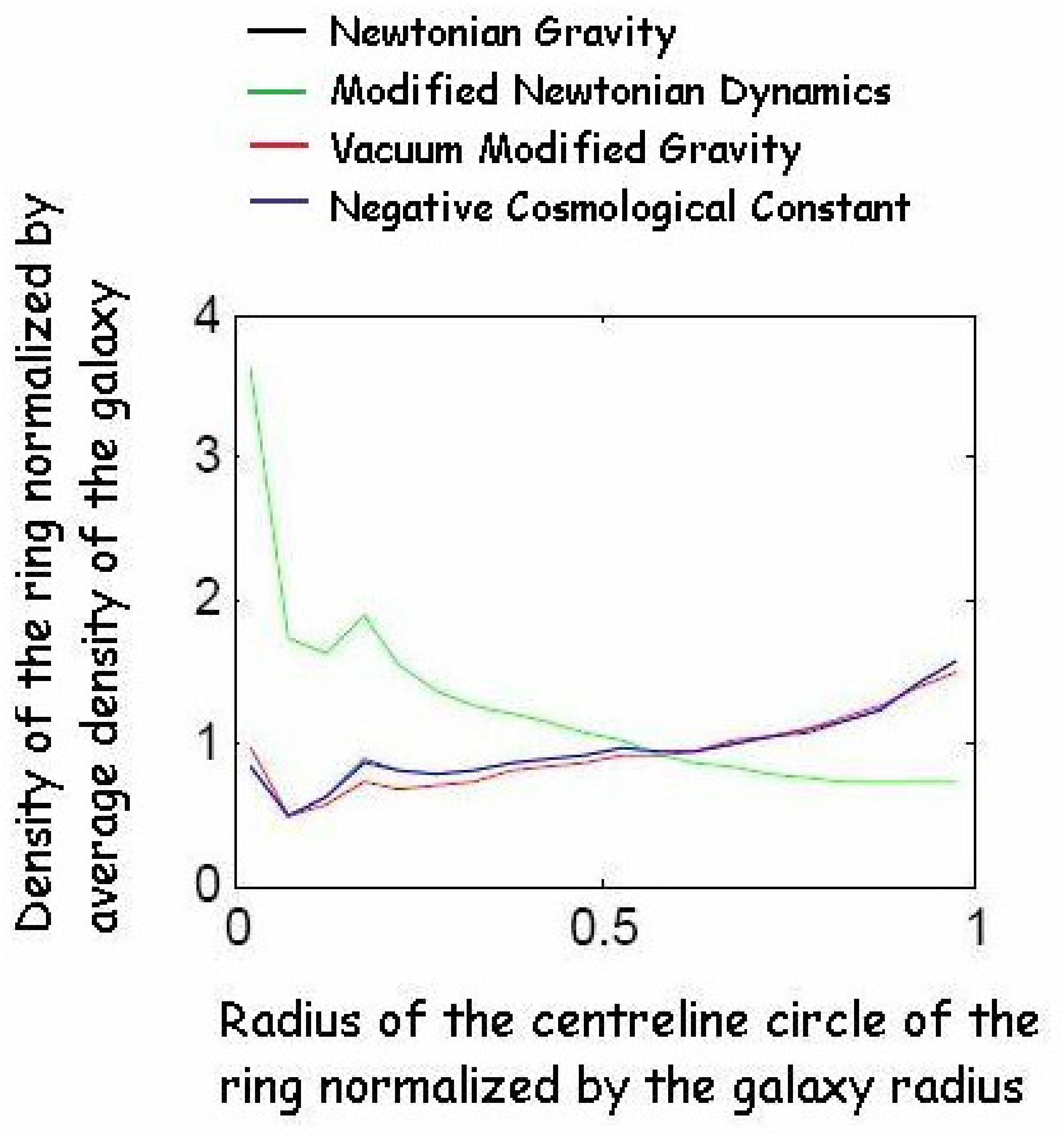
Newtonian Gravity
Modified Newtonian Dynamics
Vacuum Modified Gravity
Negative Cosmological Constant
Density of the ring normalized by average density of the galaxy
4
3
2
1
0
0
0.5
1
Radius of the centreline circle of the ring normalized by the galaxy radius

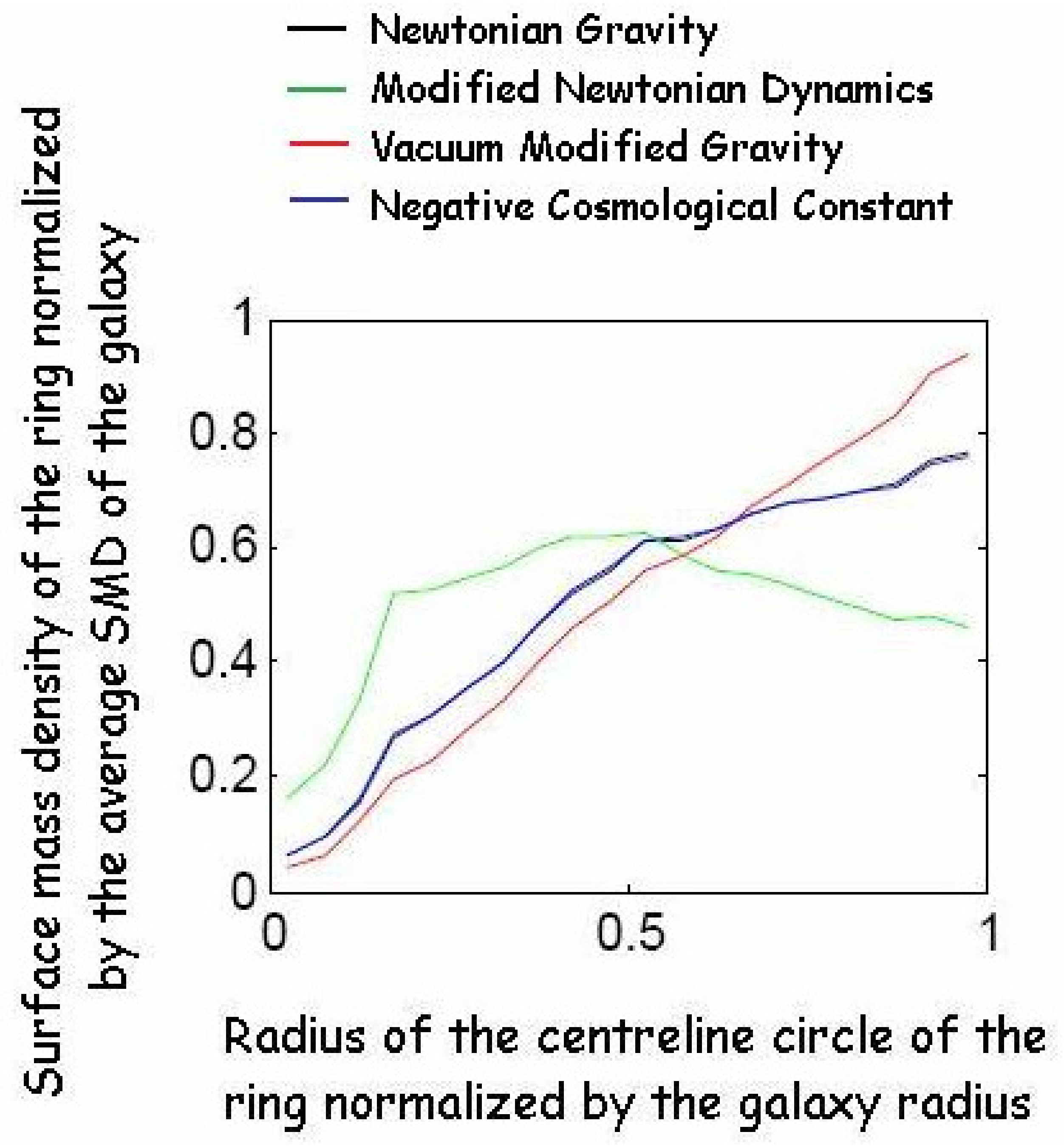

Newtonian Gravity
Modified Newtonian Dynamics
Vacuum Modified Gravity
Negative Cosmological Constant
Surface mass density of the ring normalized by the average SMD of the galaxy
1
0.8
0.6
0.4
0.2
0
0
0.5
1
Radius of the centreline circle of the ring normalized by the galaxy radius

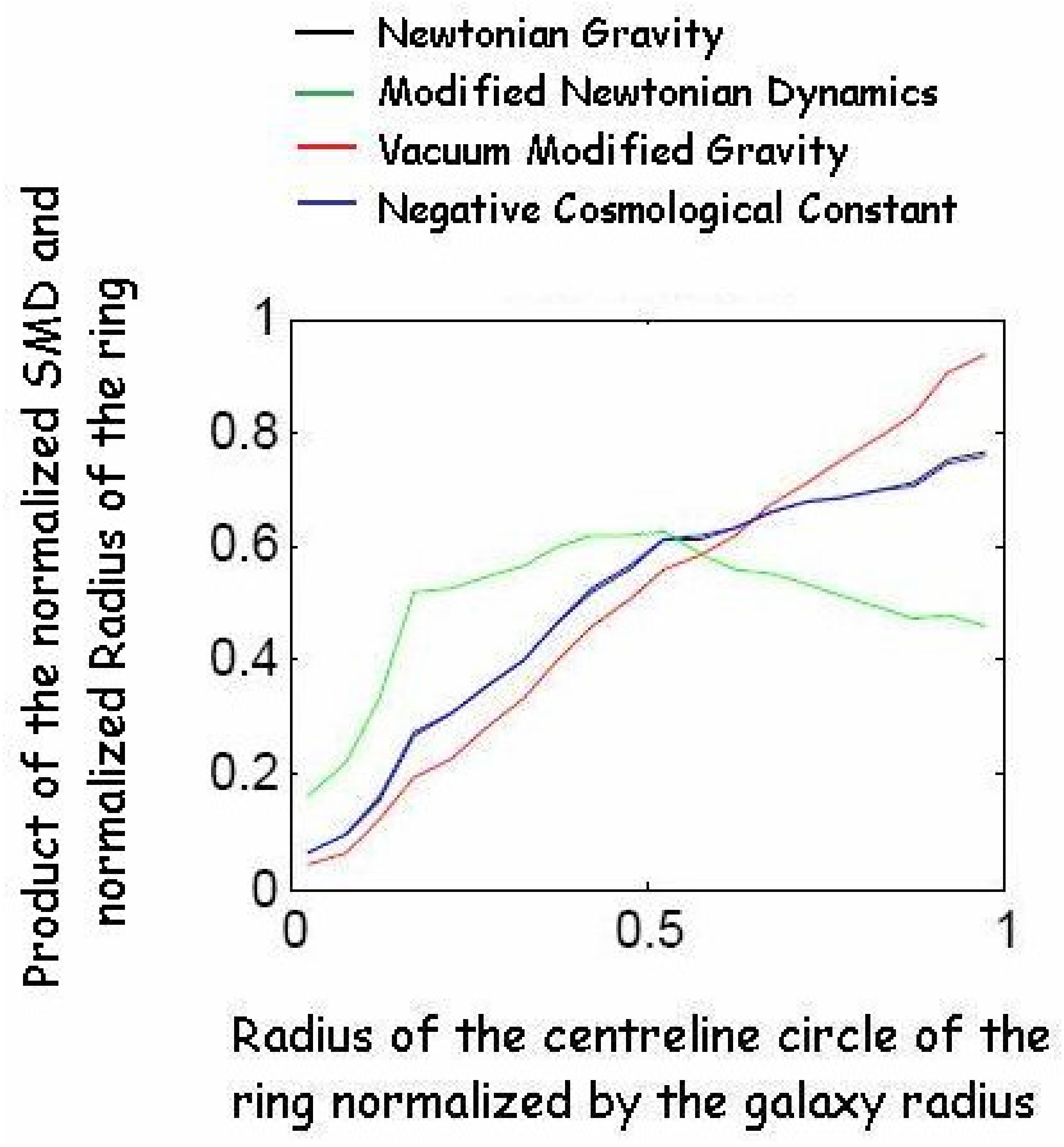
Newtonian Gravity
Modified Newtonian Dynamics
Vacuum Modified Gravity
Negative Cosmological Constant
Product of the normalized SMD and normalized Radius of the ring
1
0.8
0.6
0.4
0.2
0
0
0.5
1
Radius of the centreline circle of the ring normalized by the galaxy radius

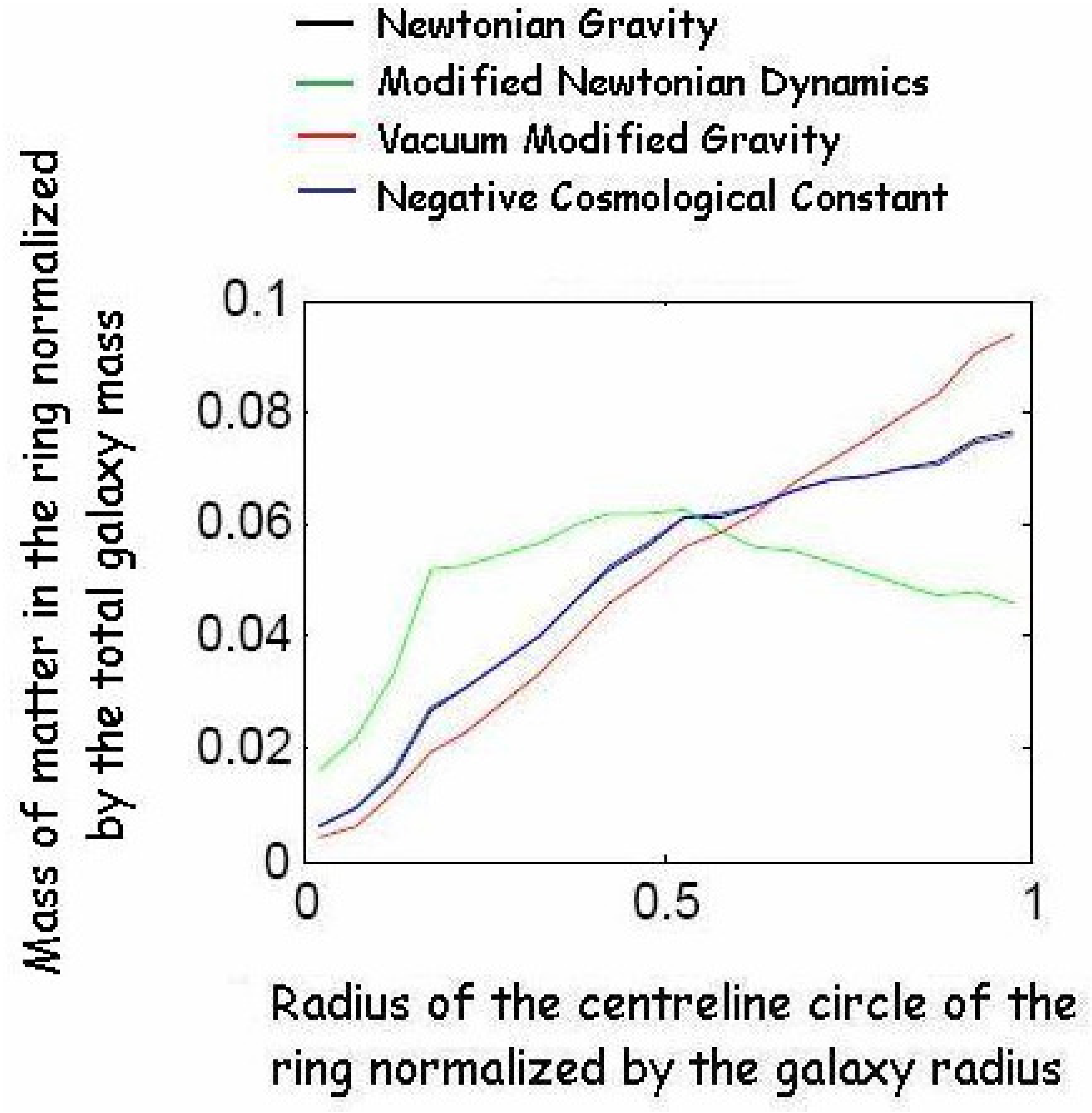
Newtonian Gravity
Modified Newtonian Dynamics
Vacuum Modified Gravity
Negative Cosmological Constant
Mass of matter in the ring normalized by the total galaxy mass
0.1
0.08
0.06
0.04
0.02
0
0
0.5
1
Radius of the centreline circle of the ring normalized by the galaxy radius

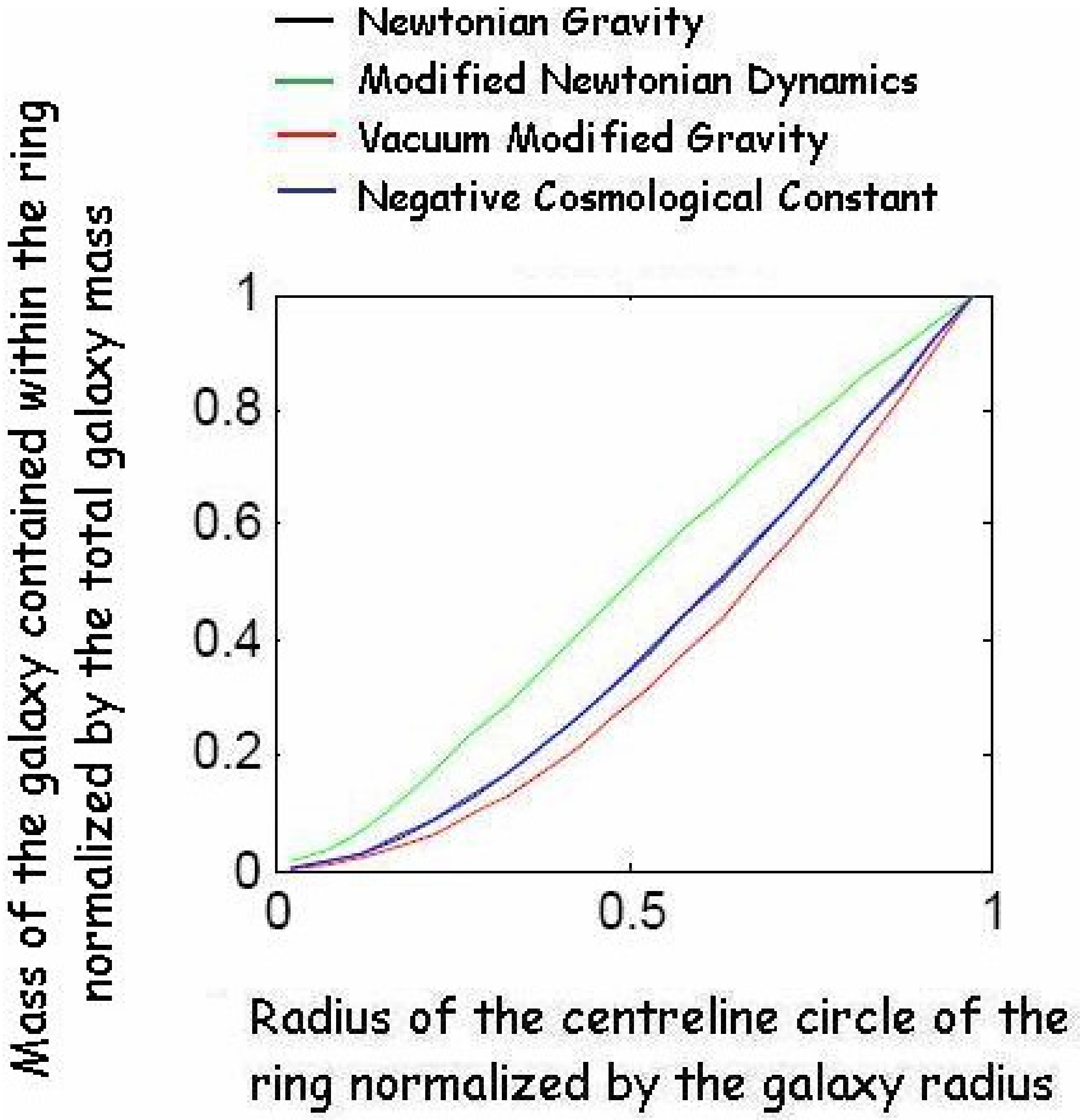
Newtonian Gravity
Modified Newtonian Dynamics
Vacuum Modified Gravity
Negative Cosmological Constant
1
0.8
0.6
0.4
0.2
0
0
0.5
1
Mass of the galaxy contained within the ring normalized by the total galaxy mass
Radius of the centreline circle of the ring normalized by the galaxy radius

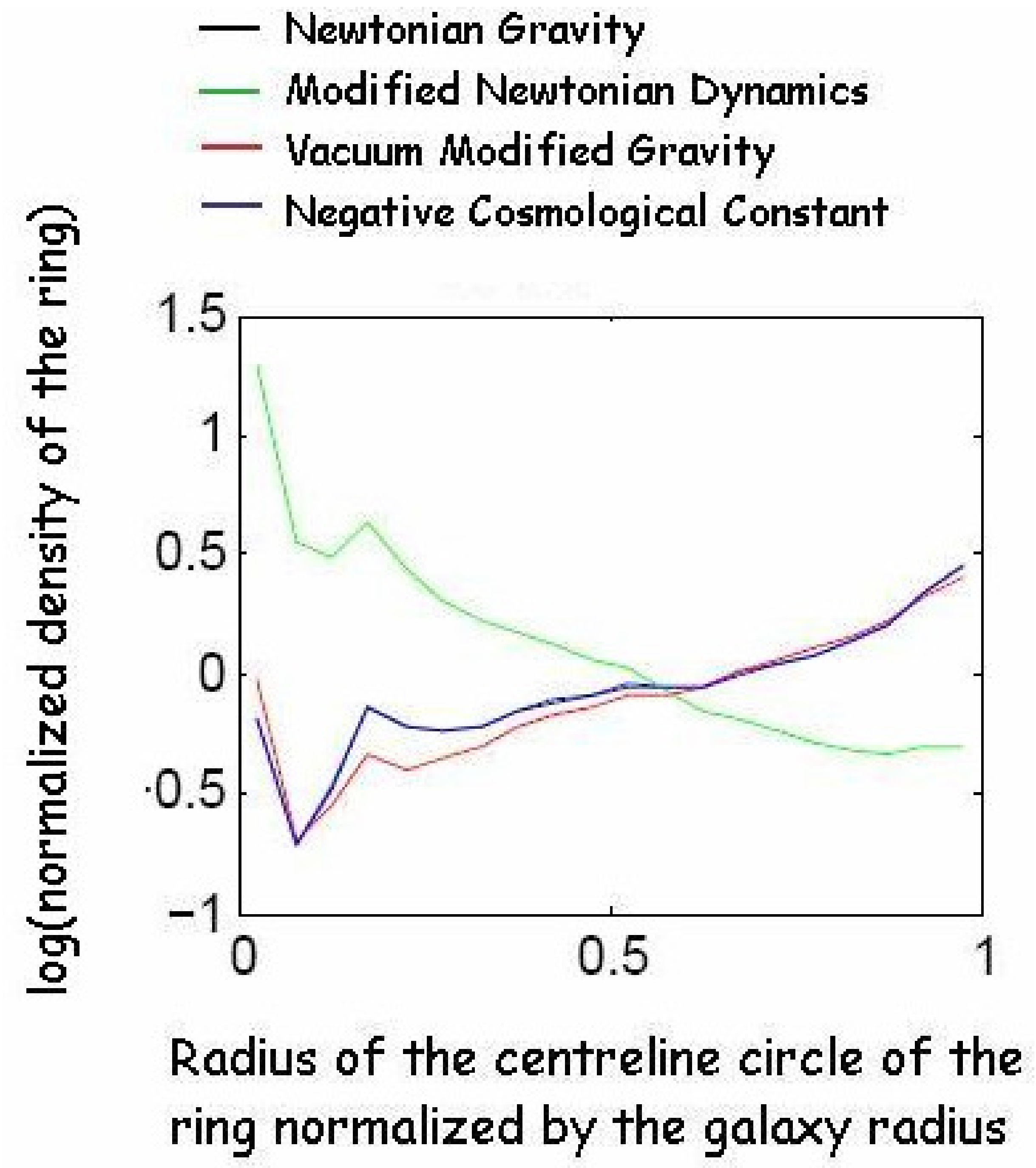
Newtonian Gravity
Modified Newtonian Dynamics
Vacuum Modified Gravity
Negative Cosmological Constant
log(normalized density of the ring)
1.5
1
0.5
0
-0.5
−1
0
0.5
1
Radius of the centreline circle of the ring normalized by the galaxy radius

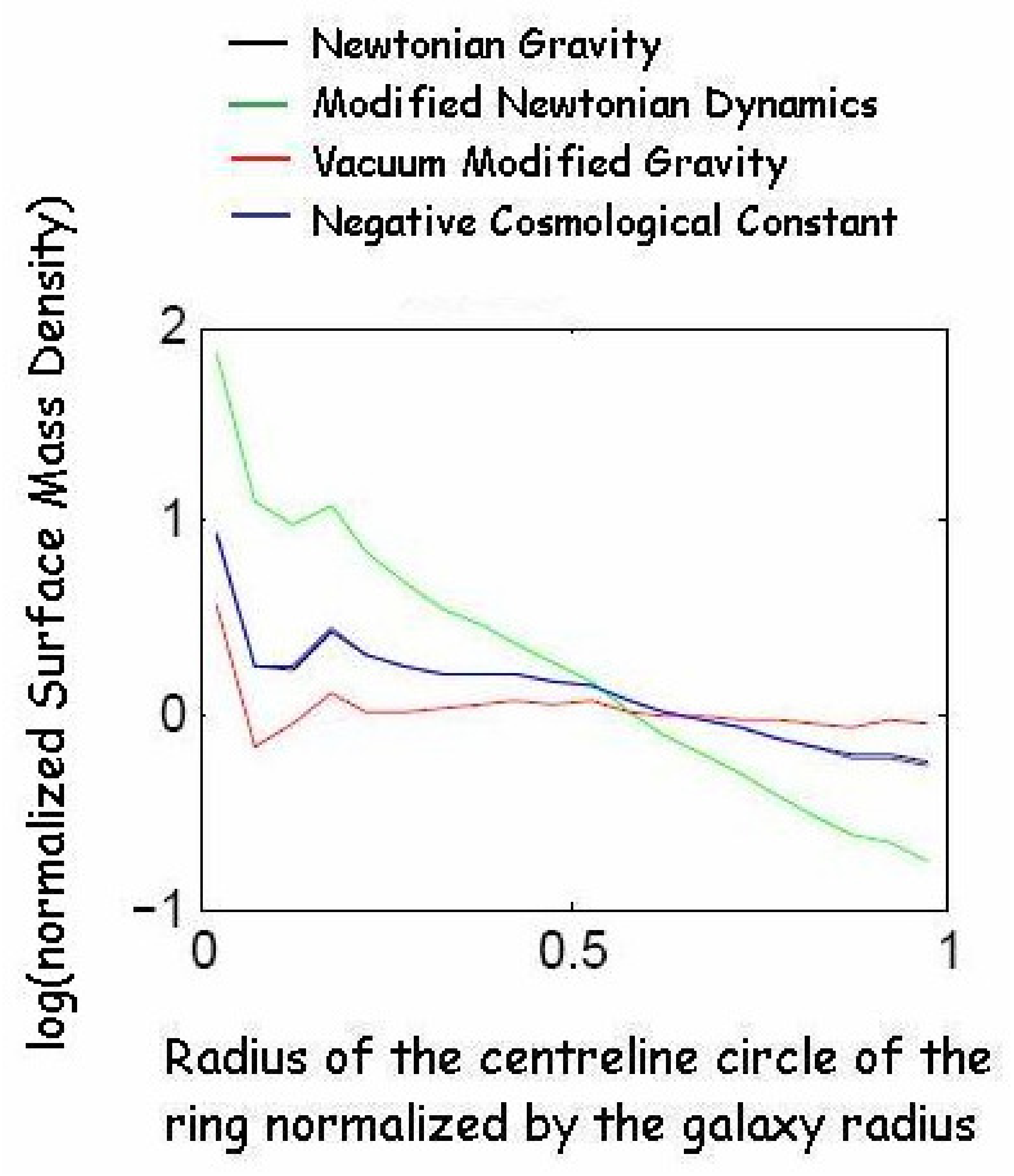
Newtonian Gravity
Modified Newtonian Dynamics
Vacuum Modified Gravity
Negative Cosmological Constant
log(normalized Surface Mass Density)
2
1
0
-1
0
0.5
1
Radius of the centreline circle of the ring normalized by the galaxy radius

## NGC 6822 (table)

| 20 Rings 4800 pc | Total Volume ($pc^3$) | Total Mass (solar mass, msuns) | Average Density ($msuns/pc^3$) | Average Surface Mass Density ($msuns/pc^2$) | Keplerian Velocity at Galaxy Rim (Kms/sec) |
|---|---|---|---|---|---|
| Newtonian Dyynamics | 7.6436 e+009 | 1.0032 E+009 | 0.1312 | 13.8593 | 29.9796 |
| Newtonian Dynamics with Negative Cosmological constant | 7.6436 e+009 | 9.8781 E+008 | 0.1292 | 13.6472 | 29.7493 |
| MoND | 1.1309 e+010 | 3.2500 E+008 | 0.0287 | 4.4901 | 17.0641 |
| Vacuum Modified Gravity | 1.1309 e+010 | 9.0237 E+008 | 0.0798 | 12.4667 | 28.4335 |

**UGC 9133 (102.5 kpc)**

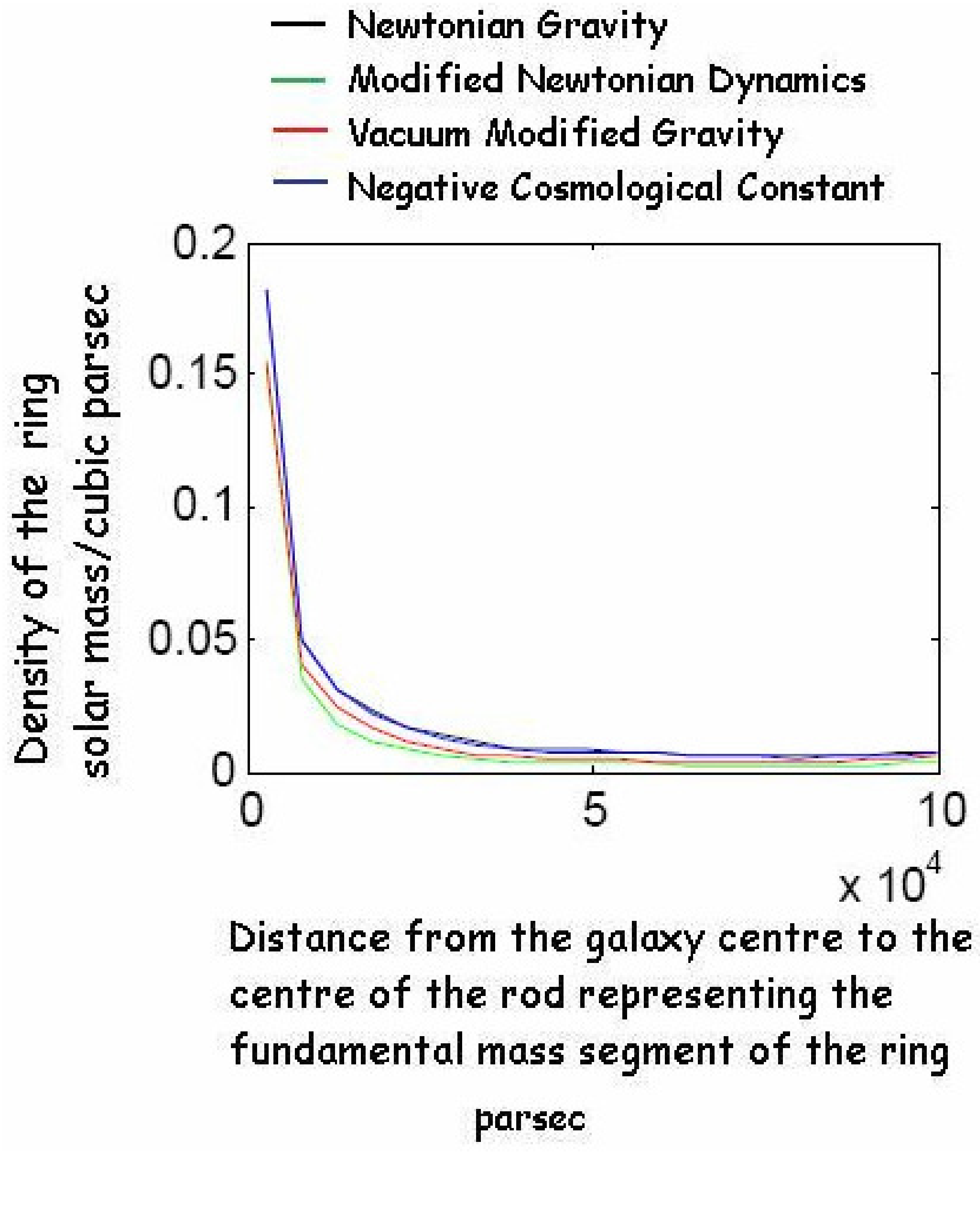

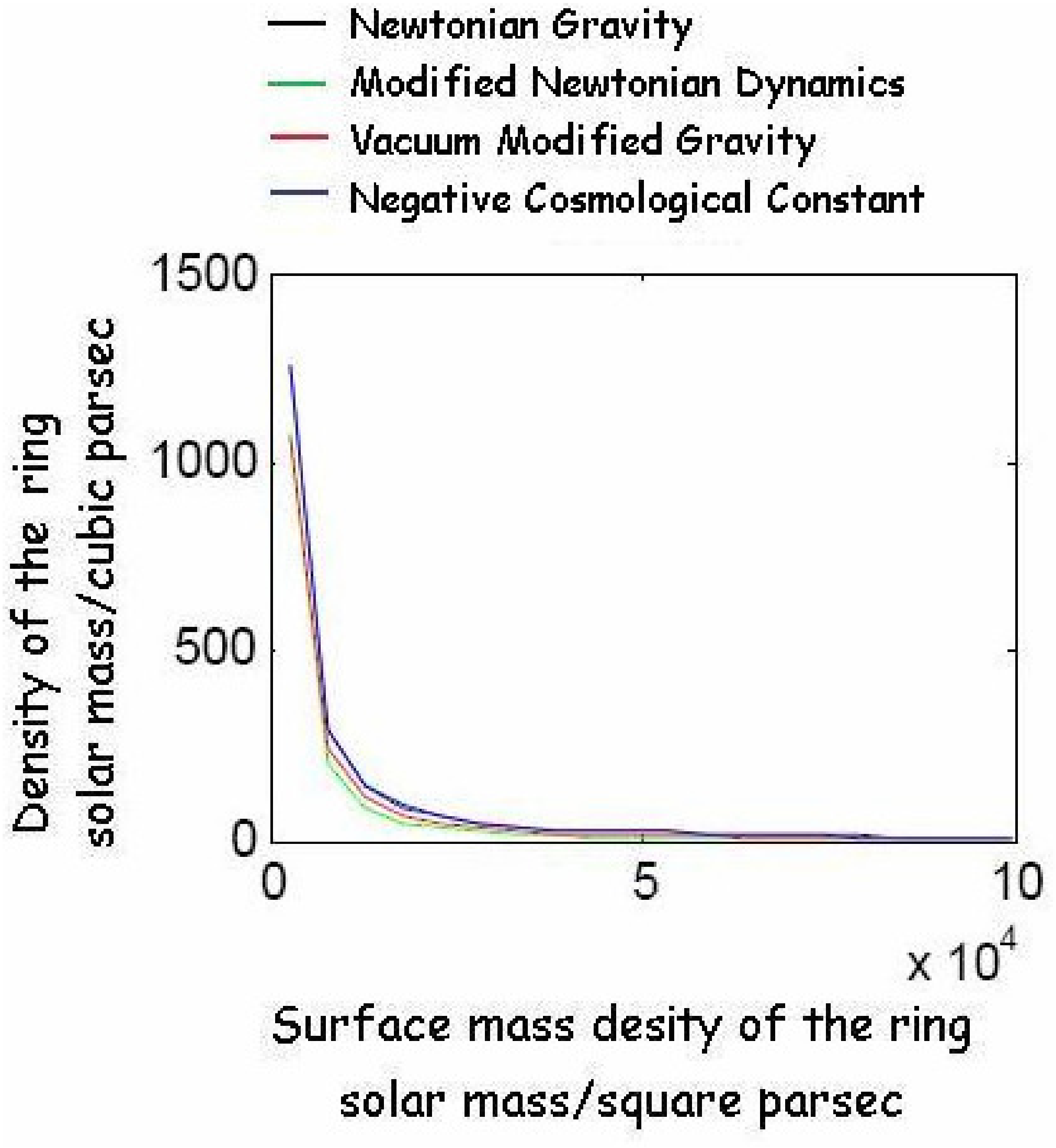

Newtonian Gravity
Modified Newtonian Dynamics
Vacuum Modified Gravity
Negative Cosmological Constant
1500
1000
500
0
0
5
10
x 10^4
Density of the ring
solar mass/cubic parsec
Surface mass desity of the ring
solar mass/square parsec

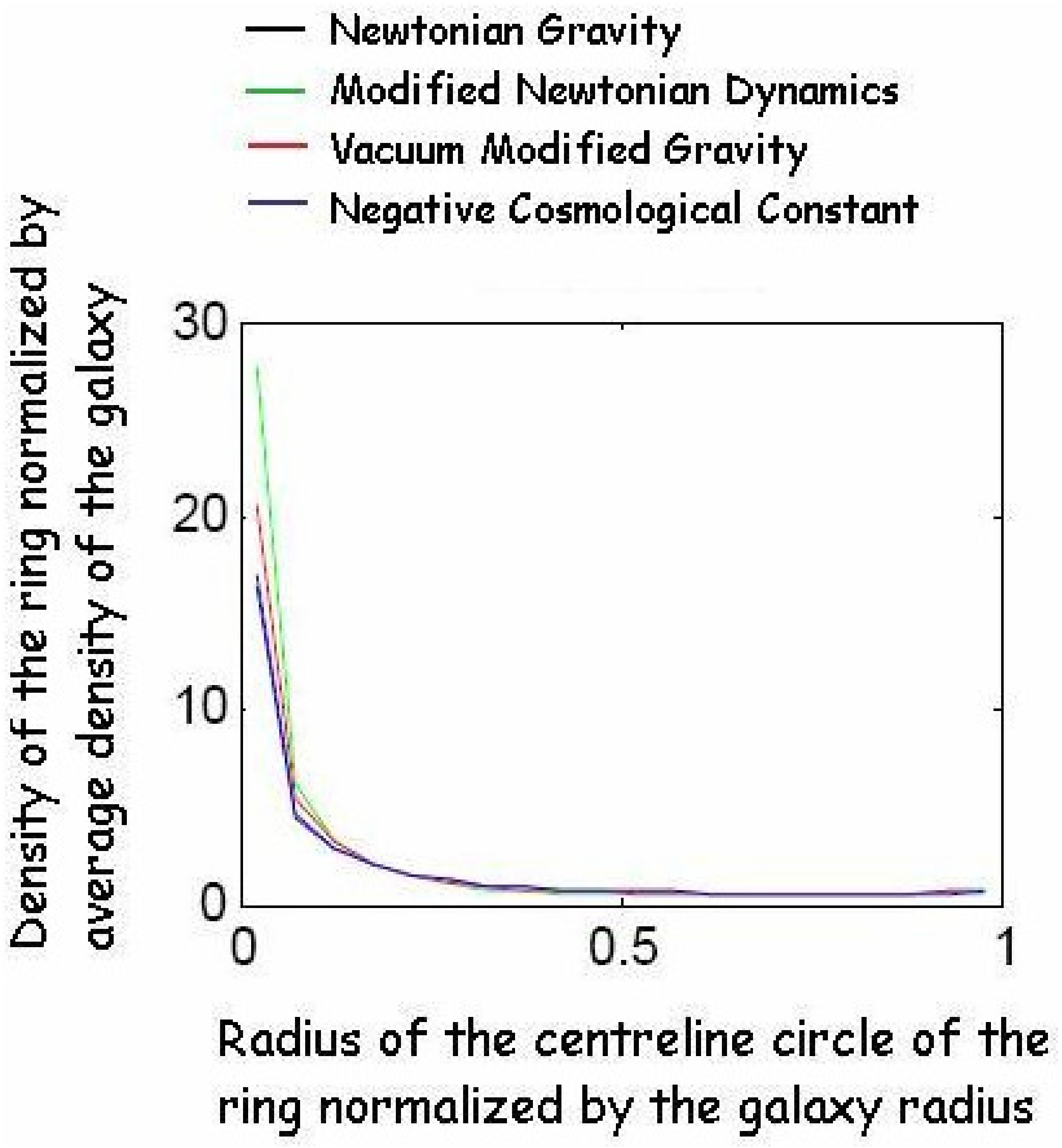
Newtonian Gravity
Modified Newtonian Dynamics
Vacuum Modified Gravity
Negative Cosmological Constant
Density of the ring normalized by average density of the galaxy
30
20
10
0
0
0.5
1
Radius of the centreline circle of the ring normalized by the galaxy radius

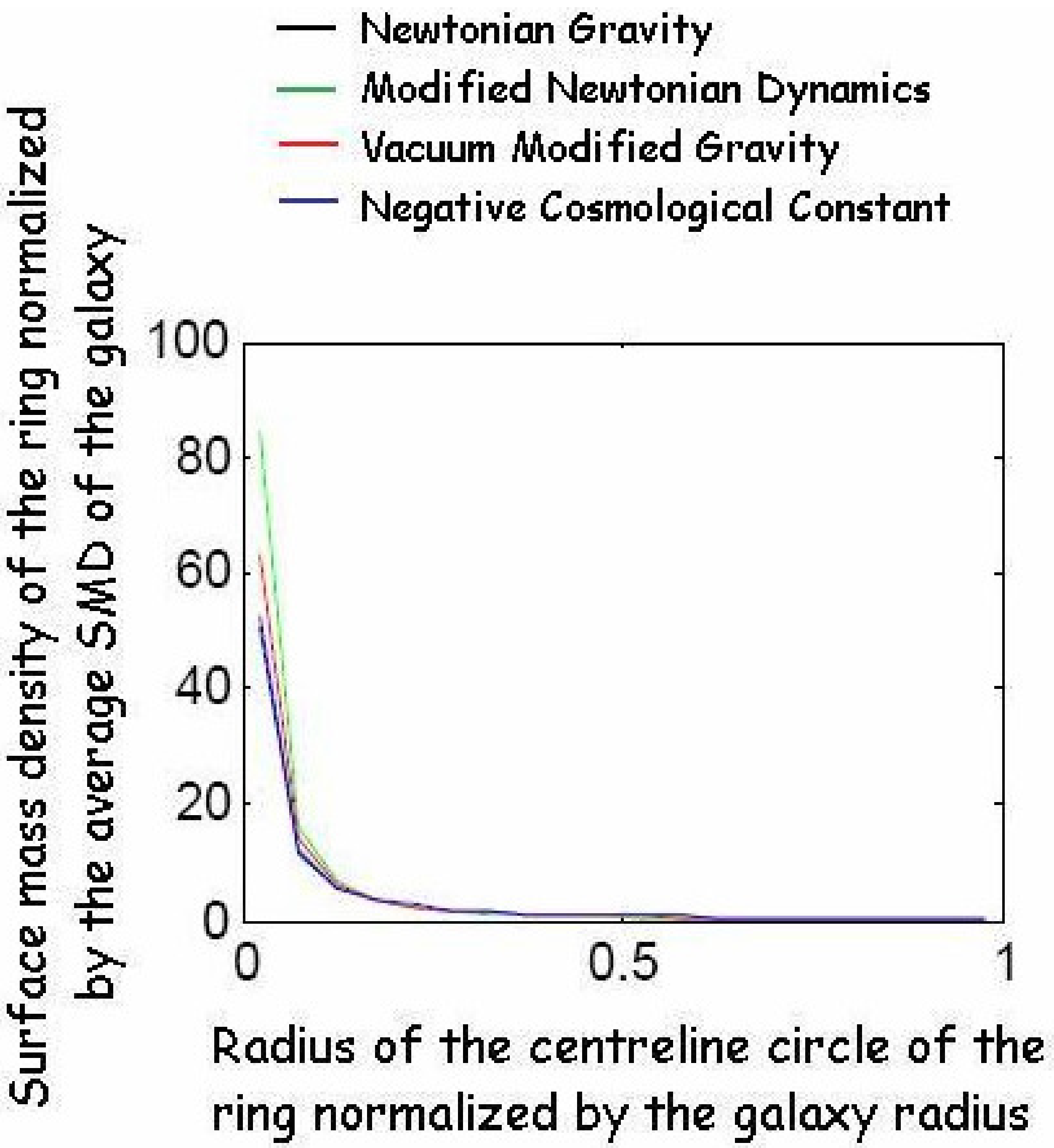
Newtonian Gravity
Modified Newtonian Dynamics
Vacuum Modified Gravity
Negative Cosmological Constant
Surface mass density of the ring normalized by the average SMD of the galaxy
100
80
60
40
20
0
0
0.5
1
Radius of the centreline circle of the ring normalized by the galaxy radius

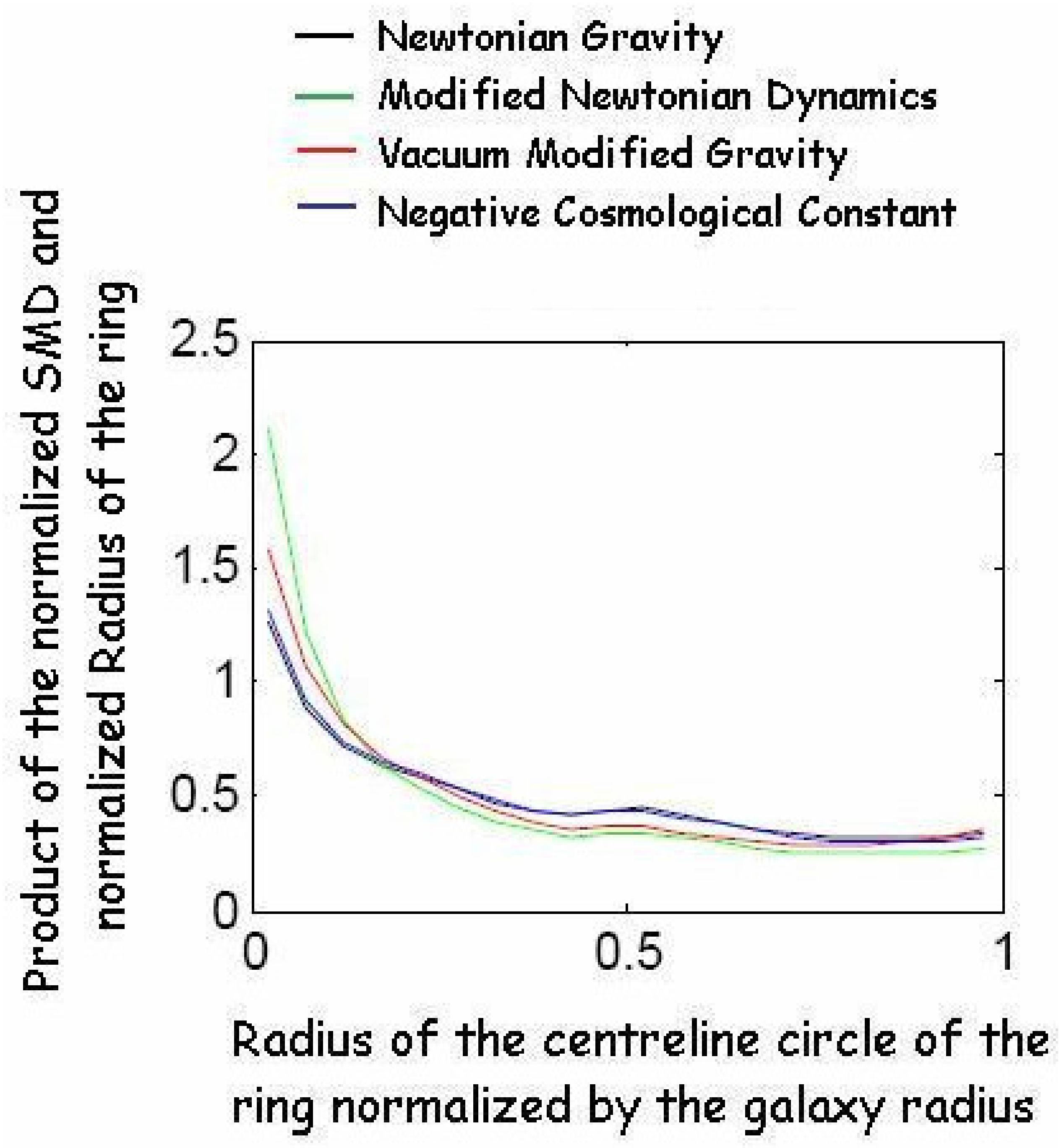
Newtonian Gravity
Modified Newtonian Dynamics
Vacuum Modified Gravity
Negative Cosmological Constant
Product of the normalized SMD and normalized Radius of the ring
2.5
2
1.5
1
0.5
0
0
0.5
1
Radius of the centreline circle of the ring normalized by the galaxy radius

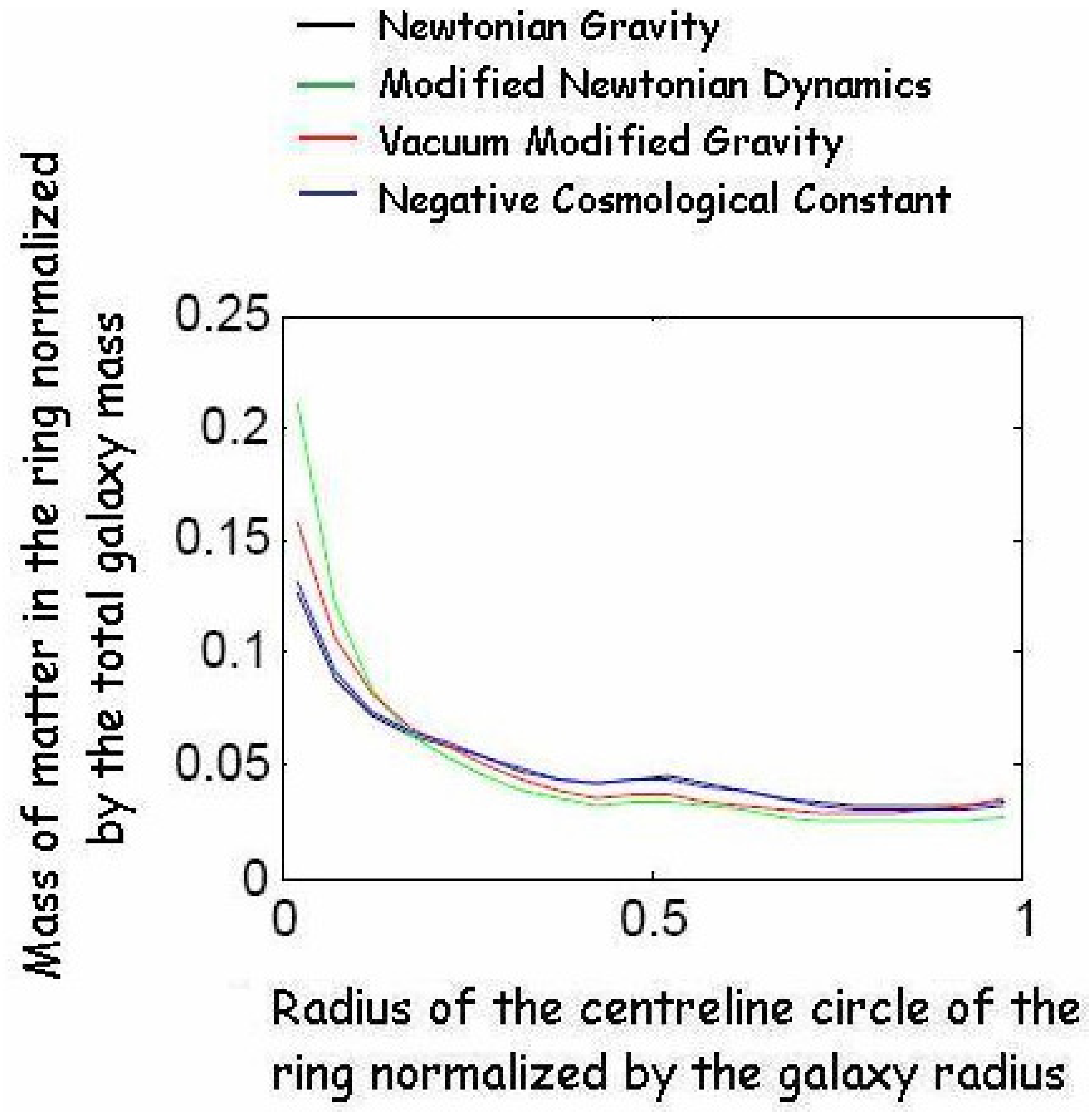
Newtonian Gravity
Modified Newtonian Dynamics
Vacuum Modified Gravity
Negative Cosmological Constant
Mass of matter in the ring normalized by the total galaxy mass
0.25
0.2
0.15
0.1
0.05
0
0
0.5
1
Radius of the centreline circle of the ring normalized by the galaxy radius

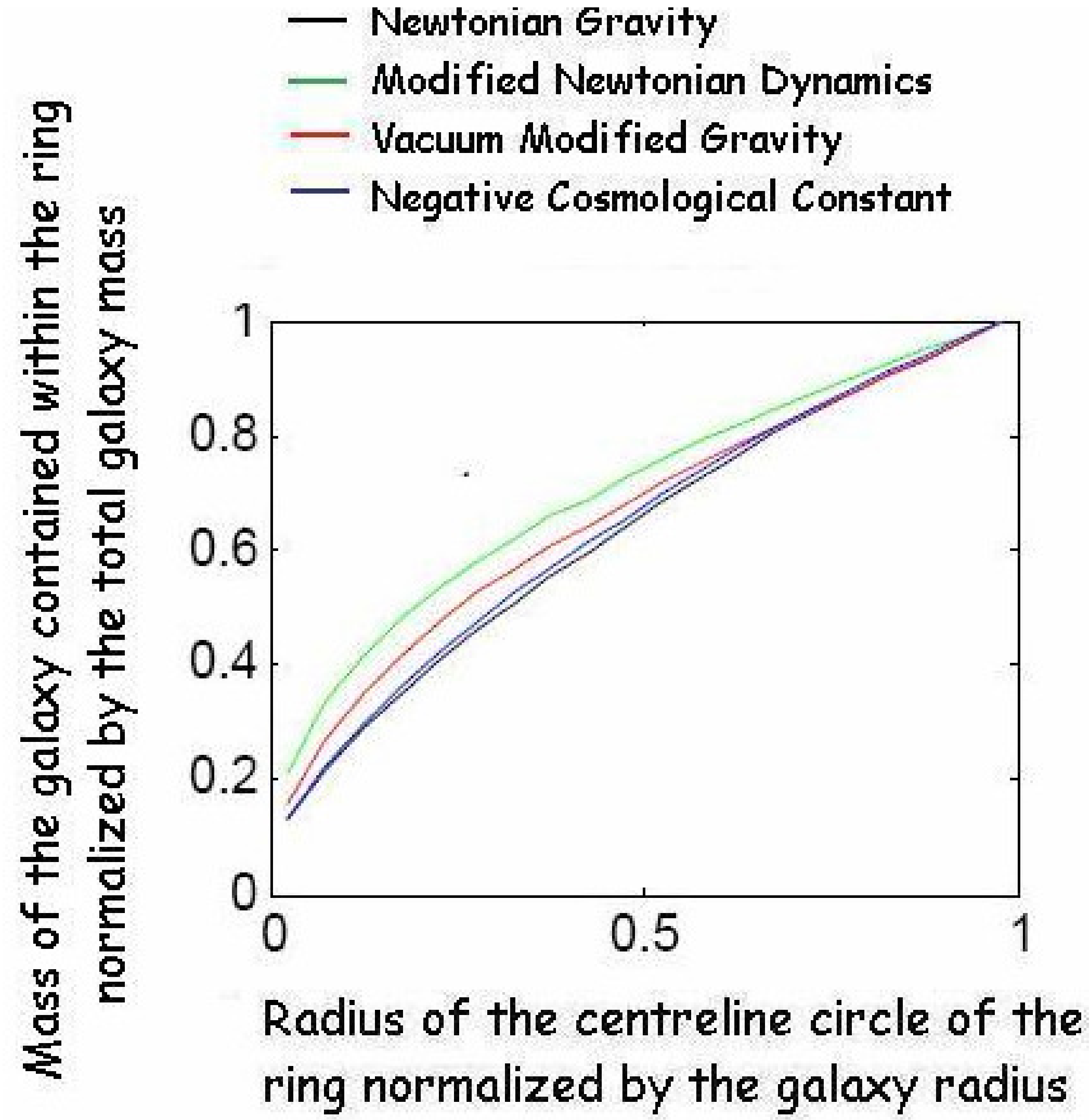

Newtonian Gravity
Modified Newtonian Dynamics
Vacuum Modified Gravity
Negative Cosmological Constant
Mass of the galaxy contained within the ring normalized by the total galaxy mass
1
0.8
0.6
0.4
0.2
0
0
0.5
1
Radius of the centreline circle of the ring normalized by the galaxy radius

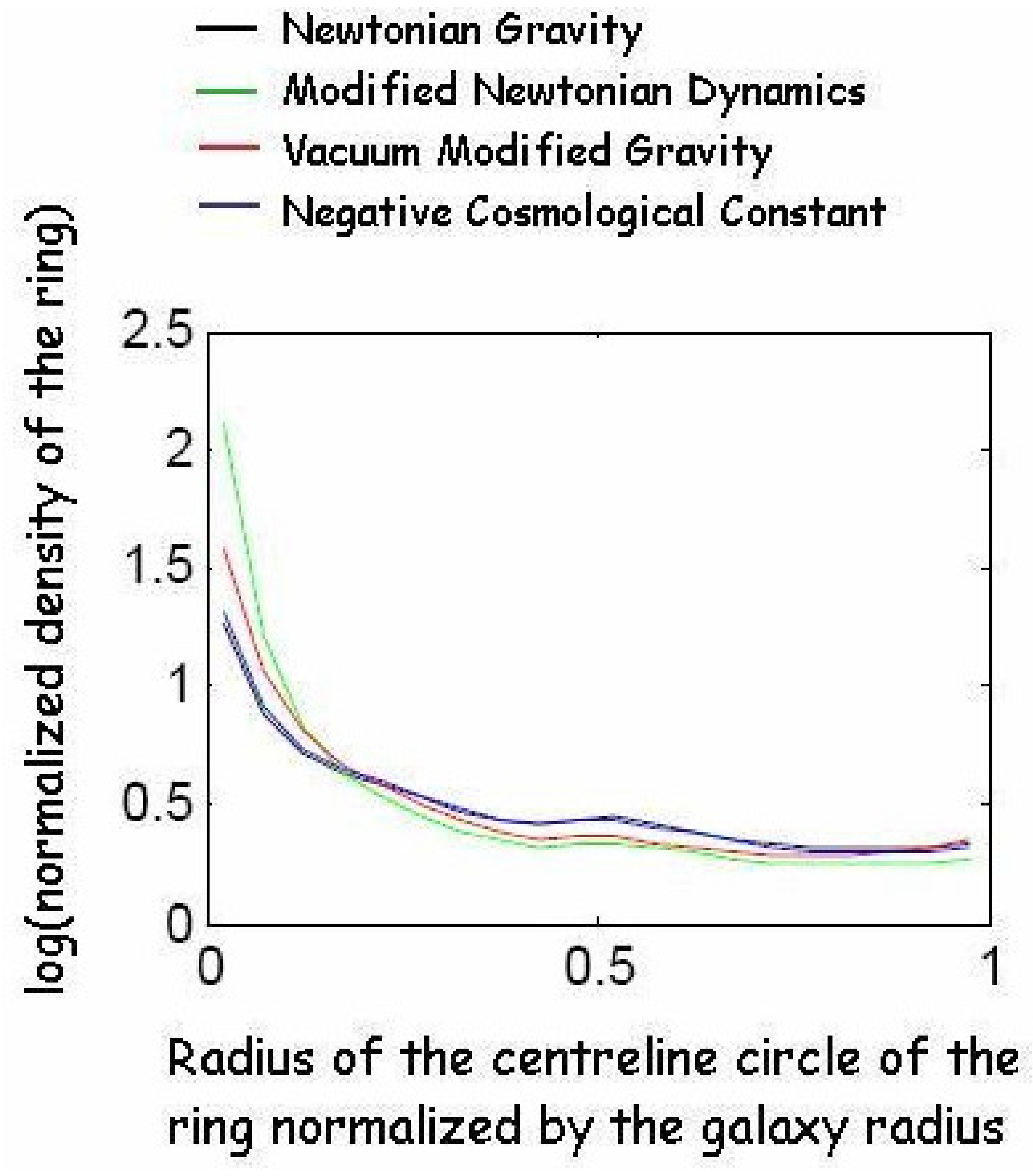
Newtonian Gravity
Modified Newtonian Dynamics
Vacuum Modified Gravity
Negative Cosmological Constant
log(normalized density of the ring)
2.5
2
1.5
1
0.5
0
0
0.5
1
Radius of the centreline circle of the ring normalized by the galaxy radius

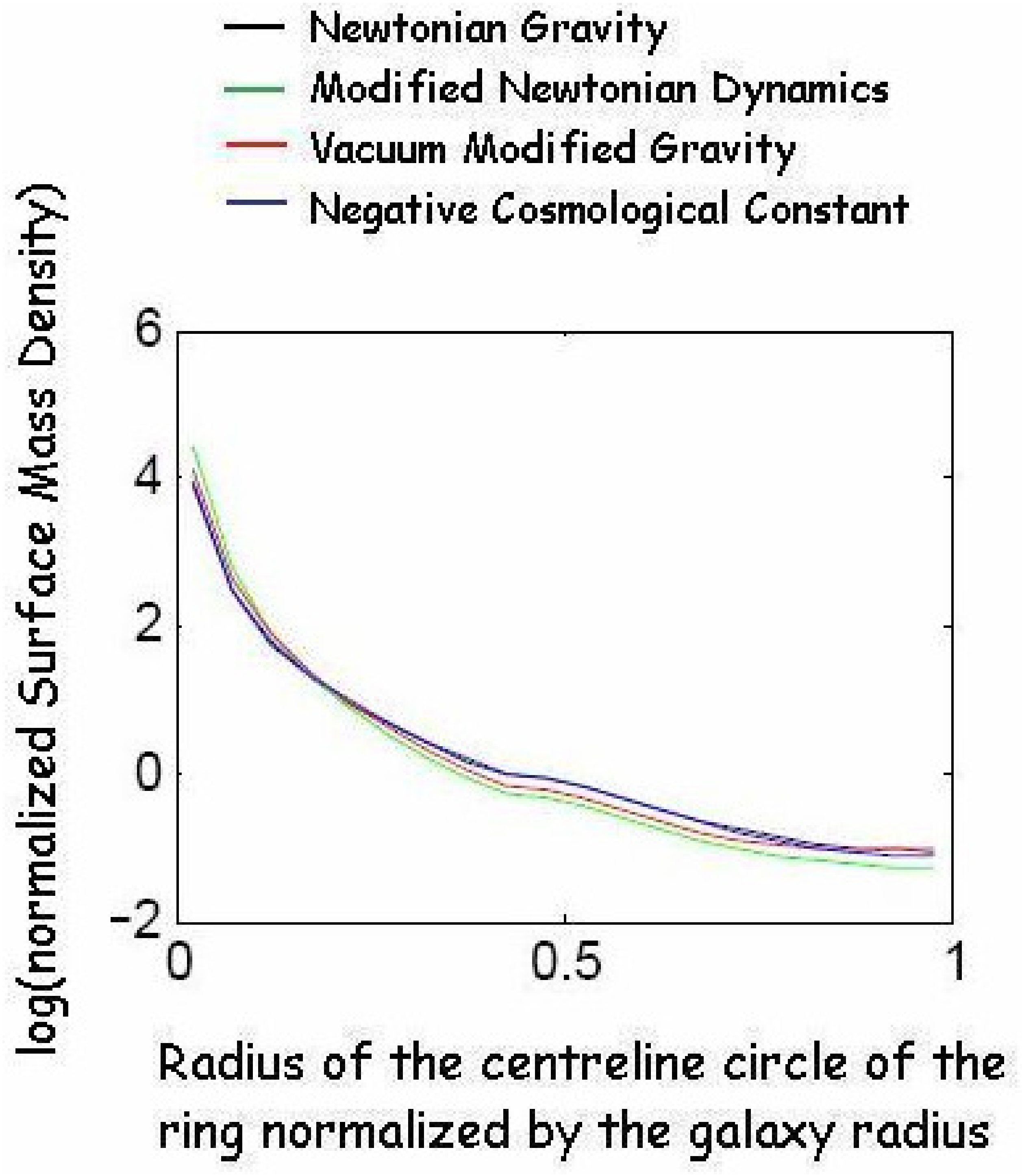
Newtonian Gravity
Modified Newtonian Dynamics
Vacuum Modified Gravity
Negative Cosmological Constant
log(normalized Surface Mass Density)
6
4
2
0
-2
0
0.5
1
Radius of the centreline circle of the ring normalized by the galaxy radius

## UGC 9133 (table)

| 20 Rings 102500 pc | Total Volume ($pc^3$) | Total Mass (solar mass, msuns) | Average Density (msuns/$pc^3$) | Average Surface Mass Density (msuns/$pc^2$) | Keplerian Velocity at Galaxy Rim (Kms/sec) |
|---|---|---|---|---|---|
| Newtonian Dyynamics | 7.4429 e+013 | 8.2719 e+011 | 0.0111 | 25.0616 | 186.2949 |
| Newtonian Dynamics with Negative Cosmological constant | 7.4429 e+013 | 7.9198 e+011 | 0.0106 | 23.9949 | 182.2871 |
| MoND | 7.4429 e+013 | 4.1810 e+011 | 0.0056 | 12.6673 | 132.4460 |
| Vacuum Modified Gravity | 7.4429 e+013 | 5.5632 e+011 | 0.0075 | 16.8548 | 152.7774 |

**Large Magellanic Cloud (LMC) (9 kpc)**

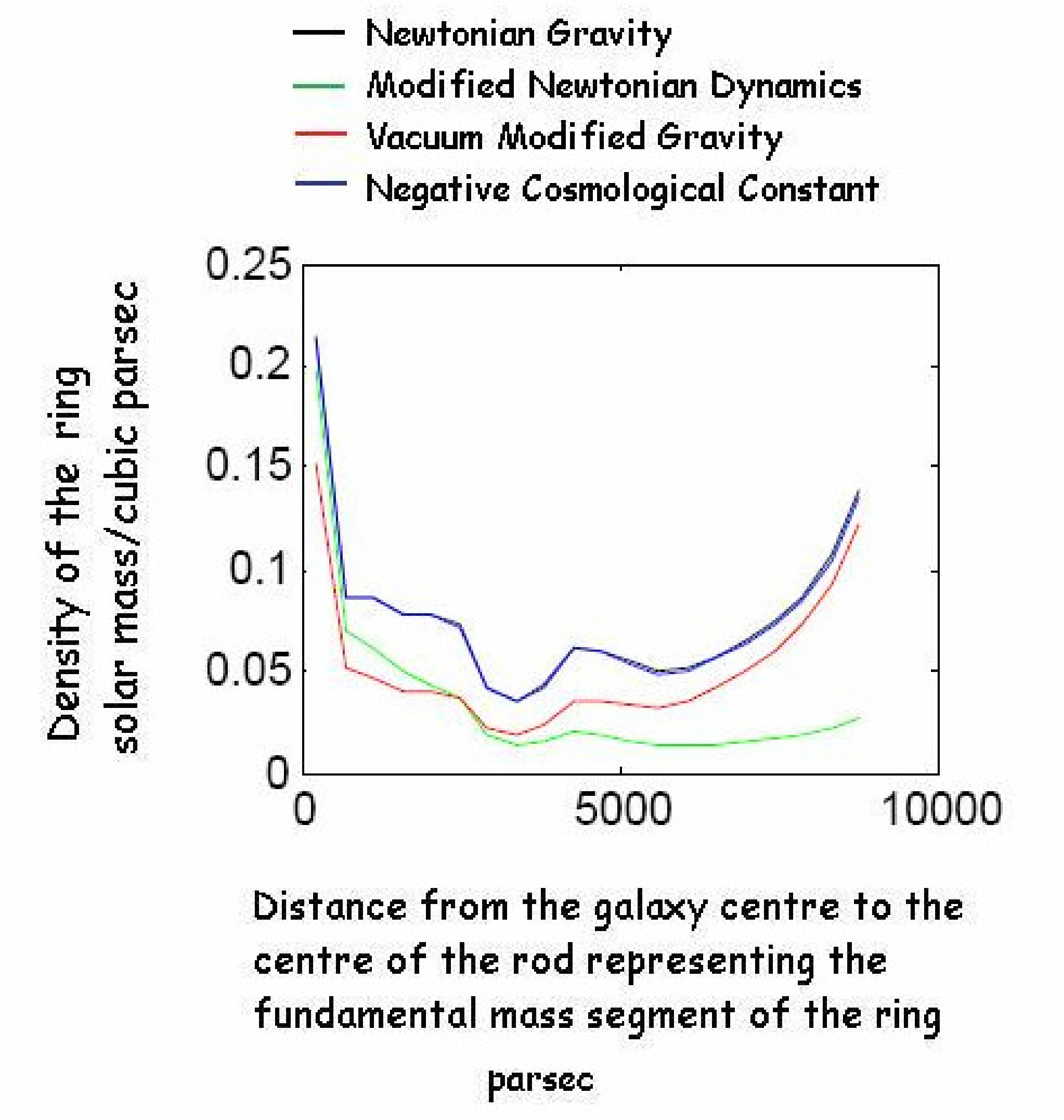

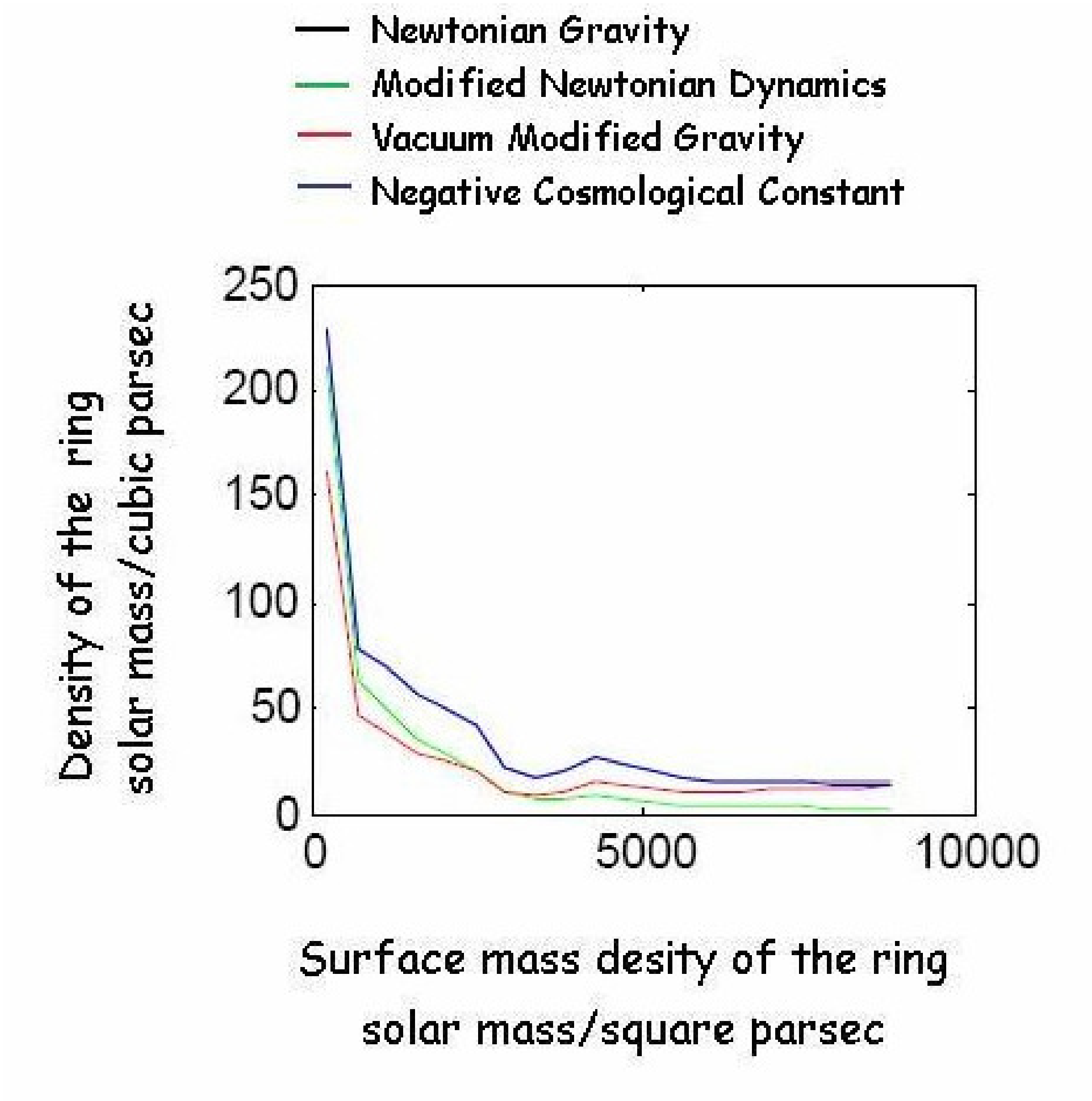
Newtonian Gravity
Modified Newtonian Dynamics
Vacuum Modified Gravity
Negative Cosmological Constant
250
200
150
100
50
0
0
5000
10000
Density of the ring
solar mass/cubic parsec
Surface mass desity of the ring
solar mass/square parsec

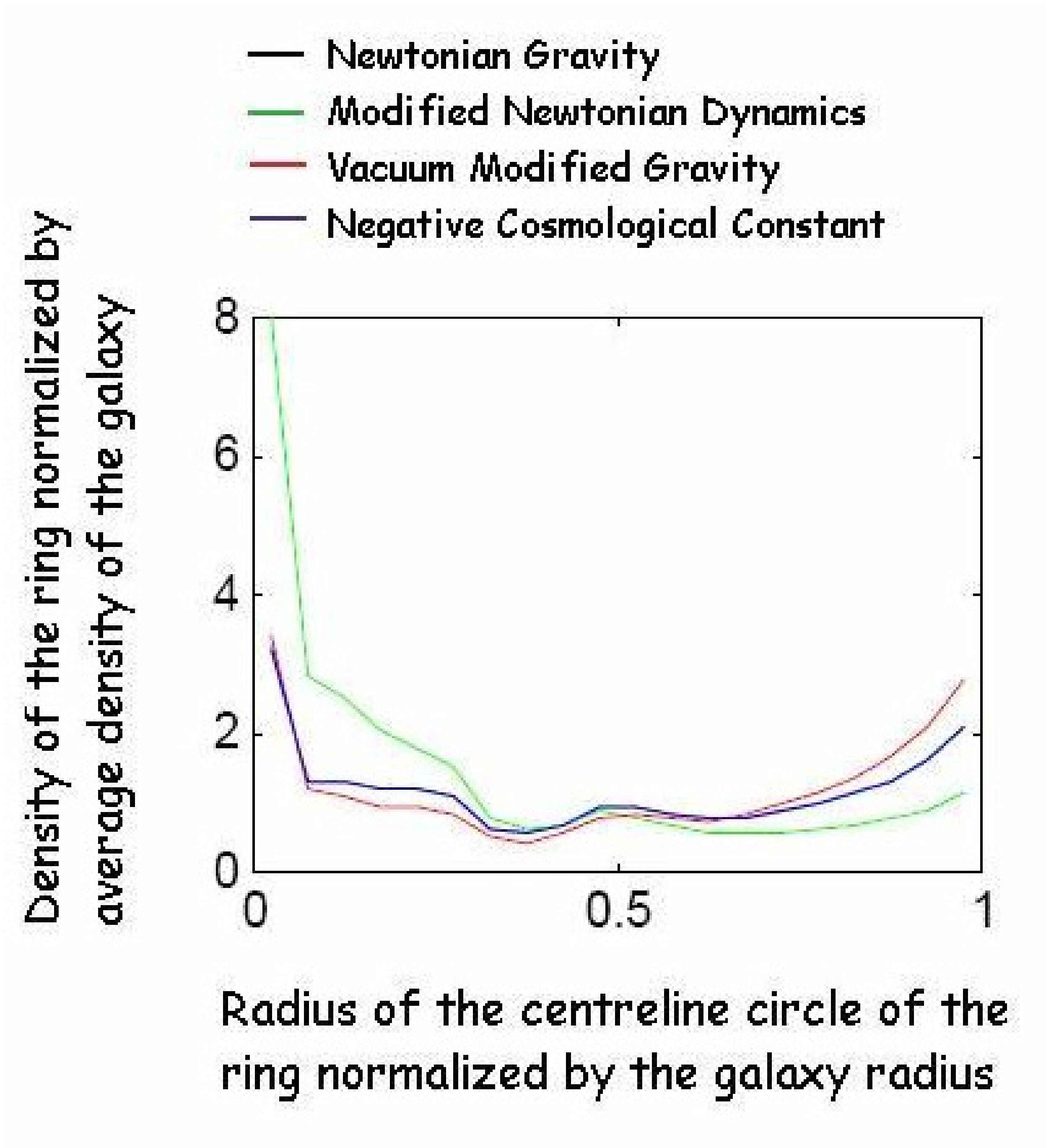

Newtonian Gravity
Modified Newtonian Dynamics
Vacuum Modified Gravity
Negative Cosmological Constant
Density of the ring normalized by average density of the galaxy
8
6
4
2
0
0
0.5
1
Radius of the centreline circle of the ring normalized by the galaxy radius

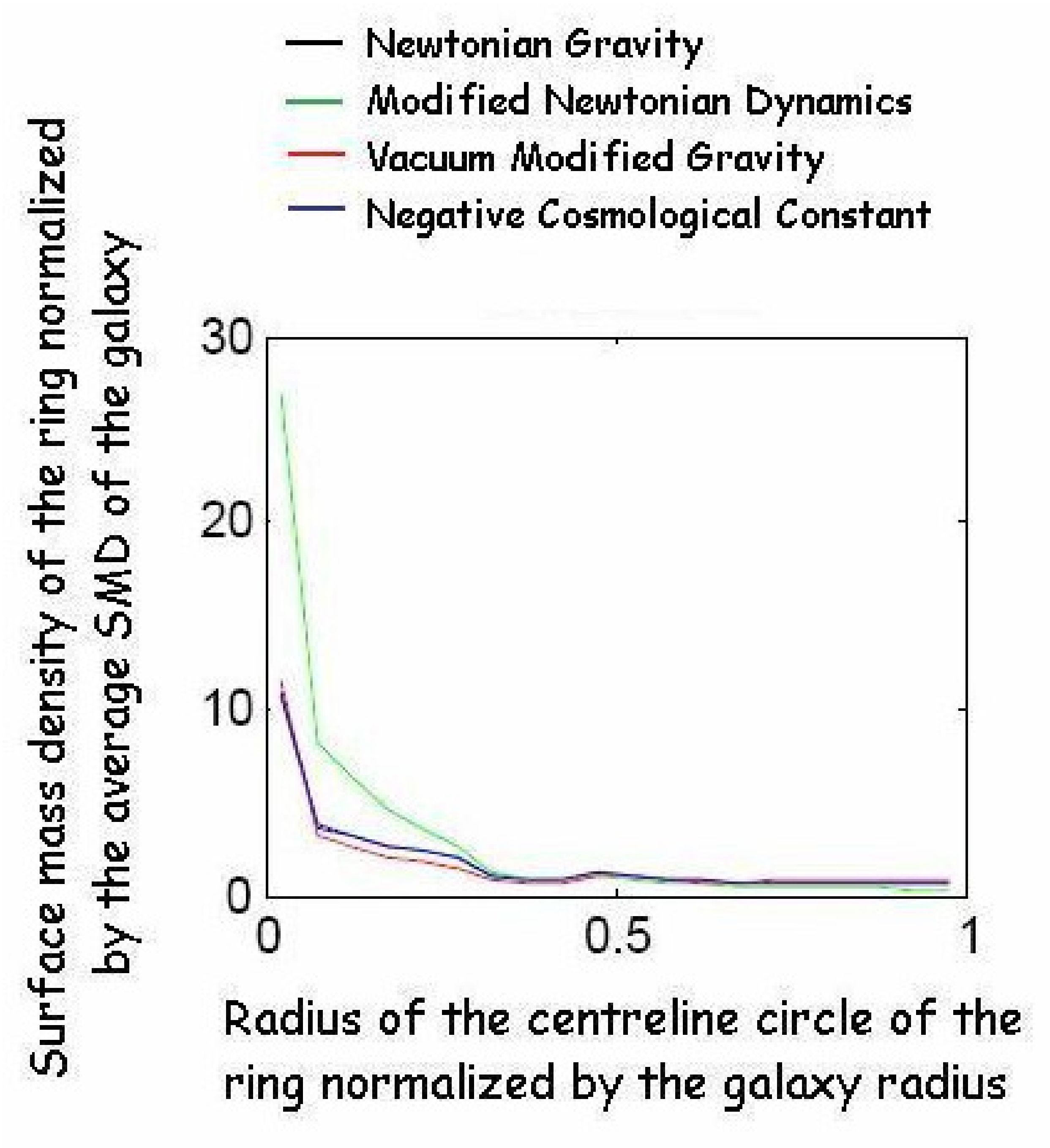
Newtonian Gravity
Modified Newtonian Dynamics
Vacuum Modified Gravity
Negative Cosmological Constant
30
20
10
0
0
0.5
1
Surface mass density of the ring normalized by the average SMD of the galaxy
Radius of the centreline circle of the ring normalized by the galaxy radius

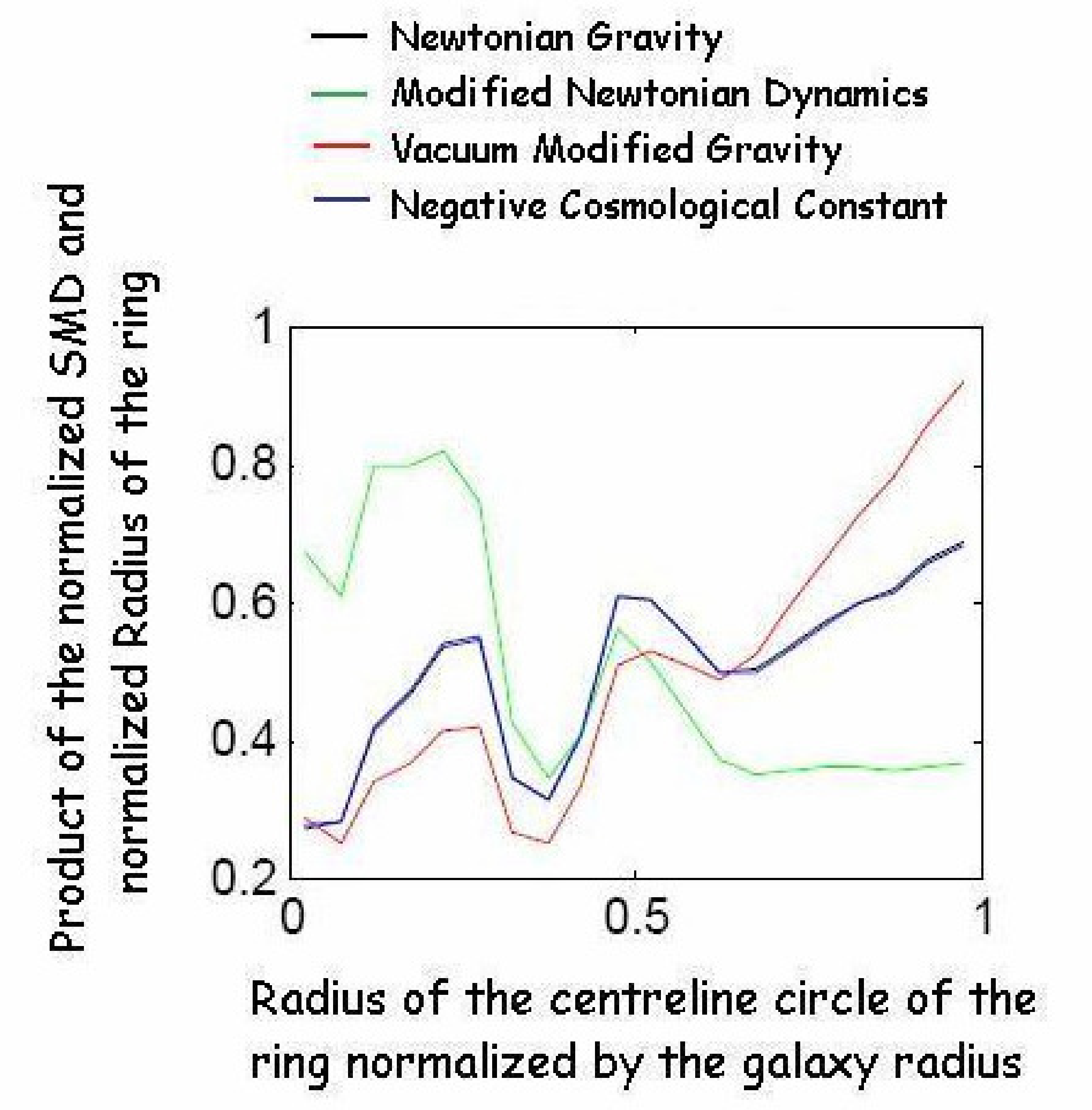

Newtonian Gravity
Modified Newtonian Dynamics
Vacuum Modified Gravity
Negative Cosmological Constant
Product of the normalized SMD and normalized Radius of the ring
1
0.8
0.6
0.4
0.2
0
0.5
1
Radius of the centreline circle of the ring normalized by the galaxy radius

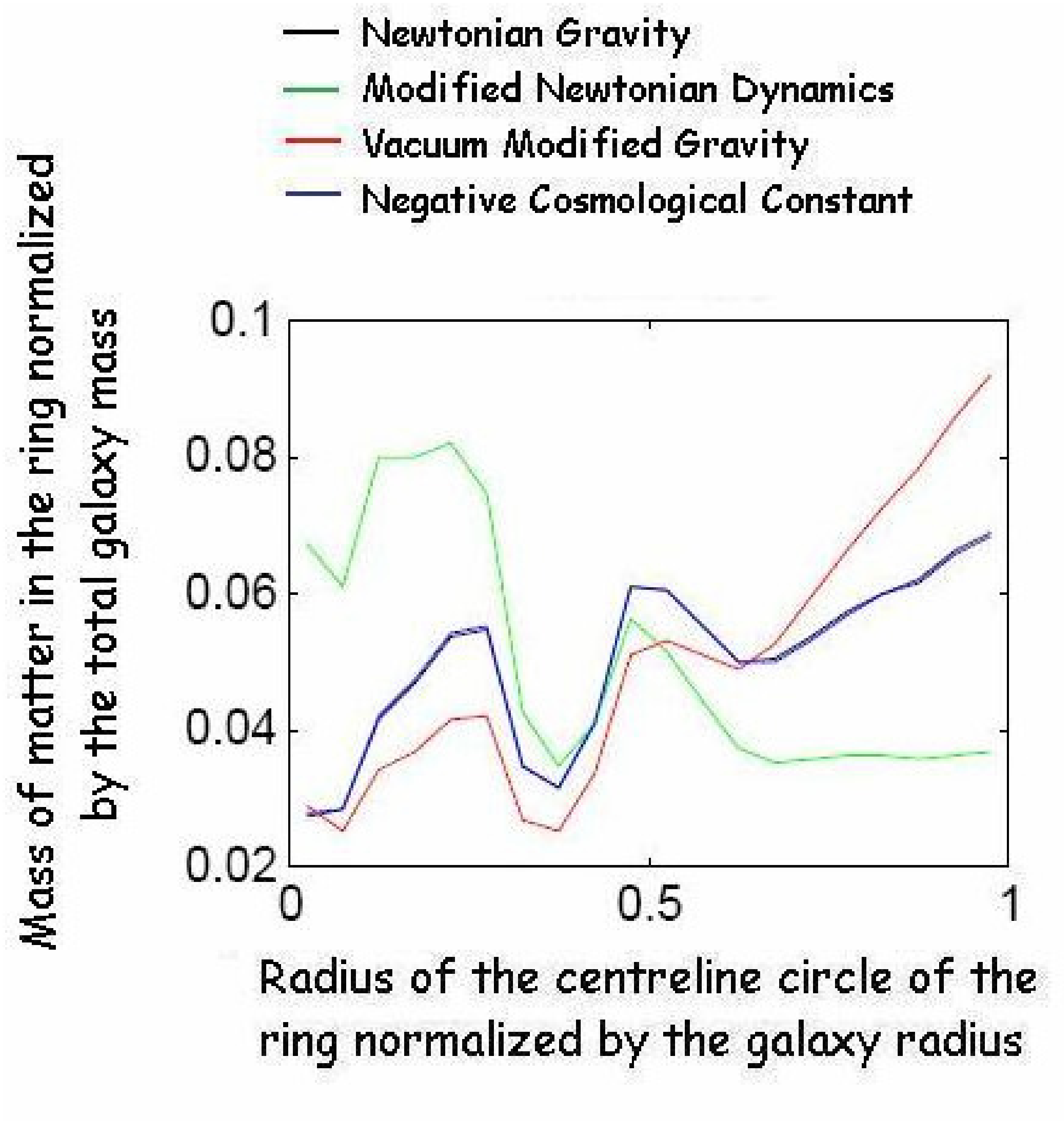
Newtonian Gravity
Modified Newtonian Dynamics
Vacuum Modified Gravity
Negative Cosmological Constant
Mass of matter in the ring normalized by the total galaxy mass
0.1
0.08
0.06
0.04
0.02
0
0.5
1
Radius of the centreline circle of the ring normalized by the galaxy radius

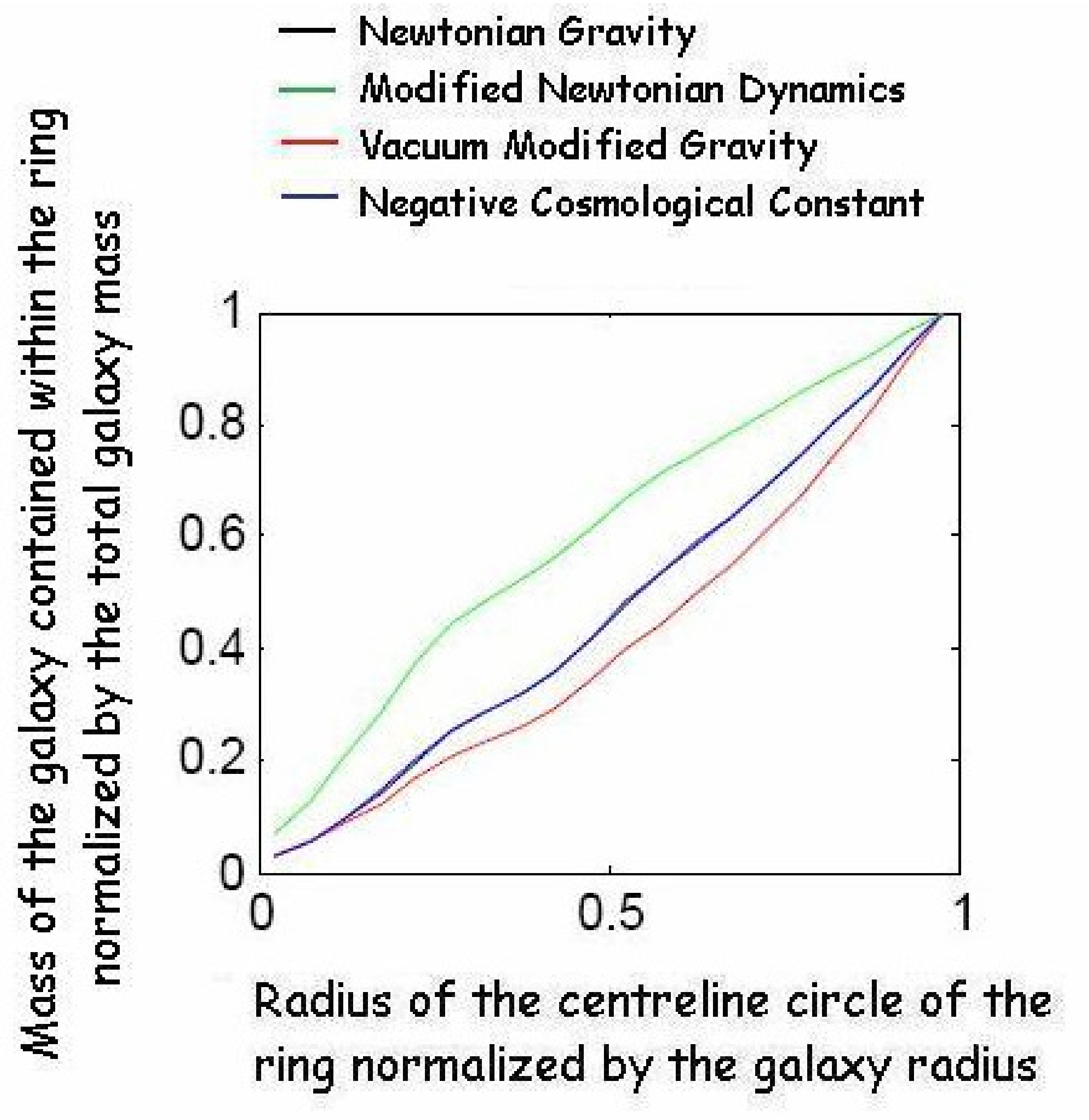
Newtonian Gravity
Modified Newtonian Dynamics
Vacuum Modified Gravity
Negative Cosmological Constant
Mass of the galaxy contained within the ring normalized by the total galaxy mass
1
0.8
0.6
0.4
0.2
0
0
0.5
1
Radius of the centreline circle of the ring normalized by the galaxy radius

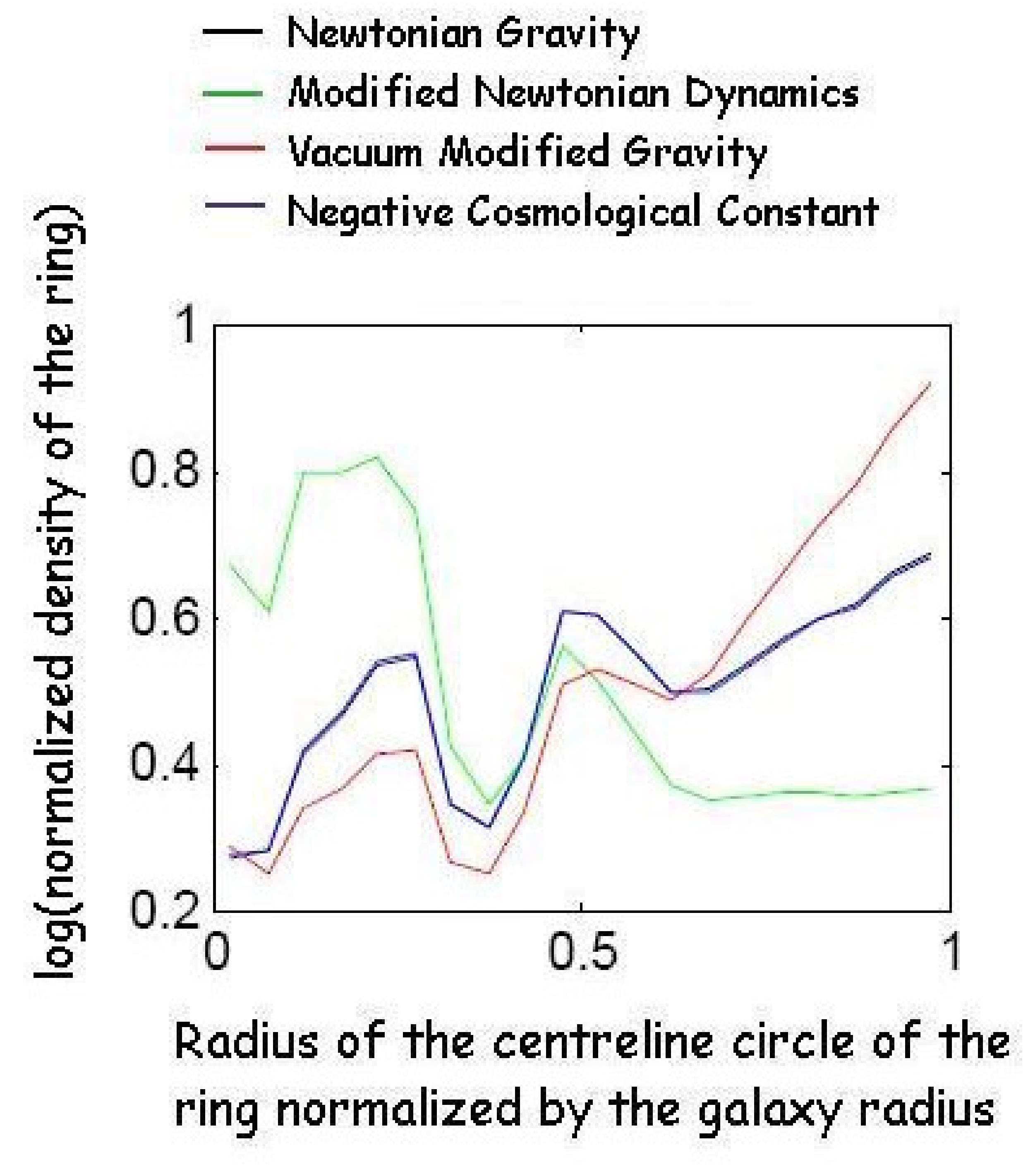

Newtonian Gravity
Modified Newtonian Dynamics
Vacuum Modified Gravity
Negative Cosmological Constant
log(normalized density of the ring)
1
0.8
0.6
0.4
0.2
0
0.5
1
Radius of the centreline circle of the ring normalized by the galaxy radius

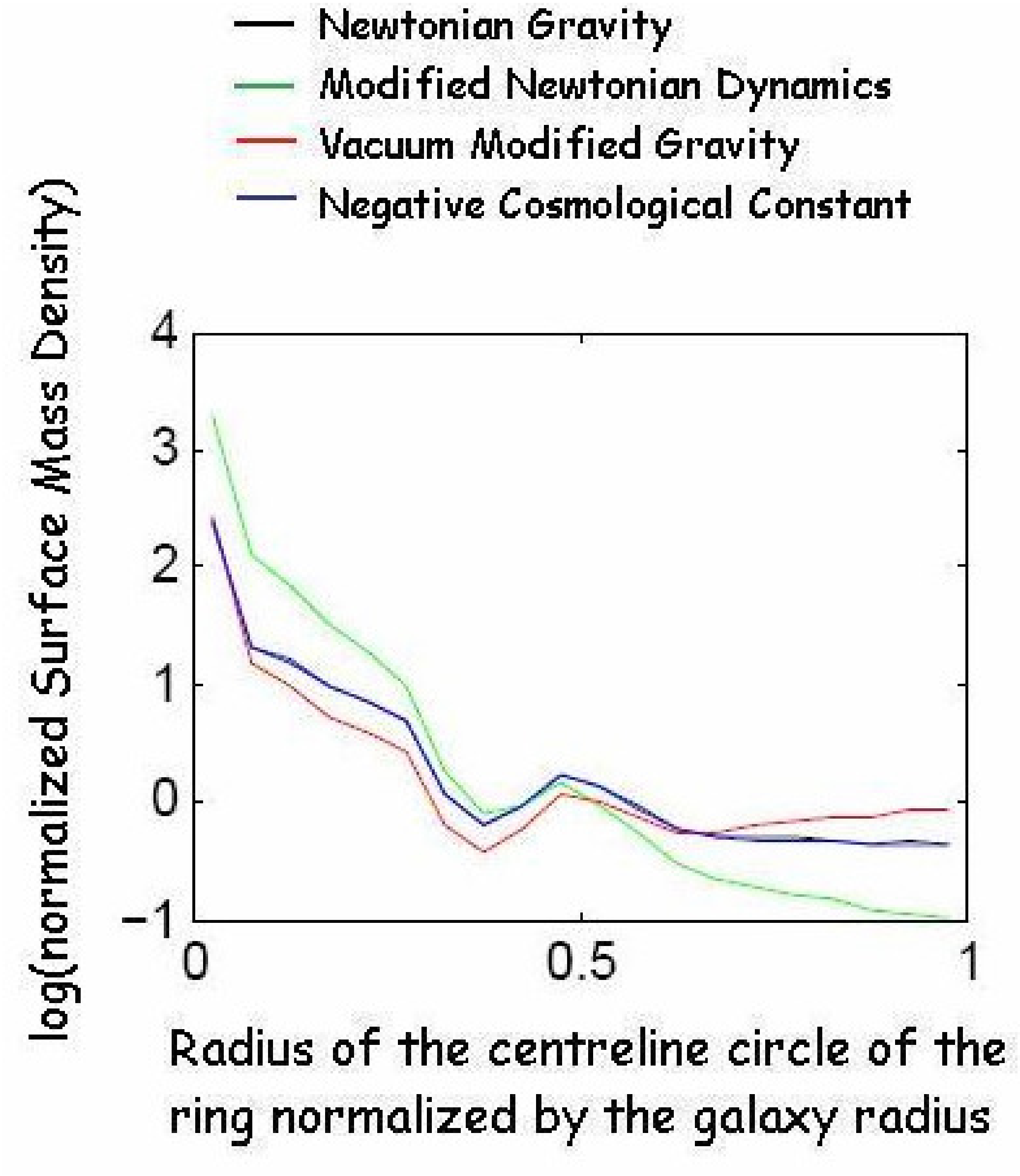
Newtonian Gravity
Modified Newtonian Dynamics
Vacuum Modified Gravity
Negative Cosmological Constant
log(normalized Surface Mass Density)
4
3
2
1
0
−1
0
0.5
1
Radius of the centreline circle of the ring normalized by the galaxy radius

## Large Magellanic Cloud (LMC) (table)

| 20 Rings 9000 pc | Total Volume ($pc^3$) | Total Mass (solar mass, msuns) | Average Density ($msuns/pc^3$) | Average Surface Mass Density ($msuns/pc^2$) | Keplerian Velocity at Galaxy Rim (Kms/sec) |
|---|---|---|---|---|---|
| Newtonian Dyynamics | 8.0439 e+010 | 5.3772 E+009 | 0.0668 | 21.1309 | 50.6891 |
| Newtonian Dynamics with Negative Cosmological constant | 8.0439 e+010 | 5.2885 E+009 | 0.0657 | 20.7824 | 50.2694 |
| MoND | 8.0439 e+010 | 1.9989 E+009 | 0.0248 | 7.8551 | 30.9052 |
| Vacuum Modified Gravity | 8.0439 e+010 | 3.5654 E+009 | 0.0443 | 14.0113 | 41.2758 |

**Milky Way (17 kpc)**

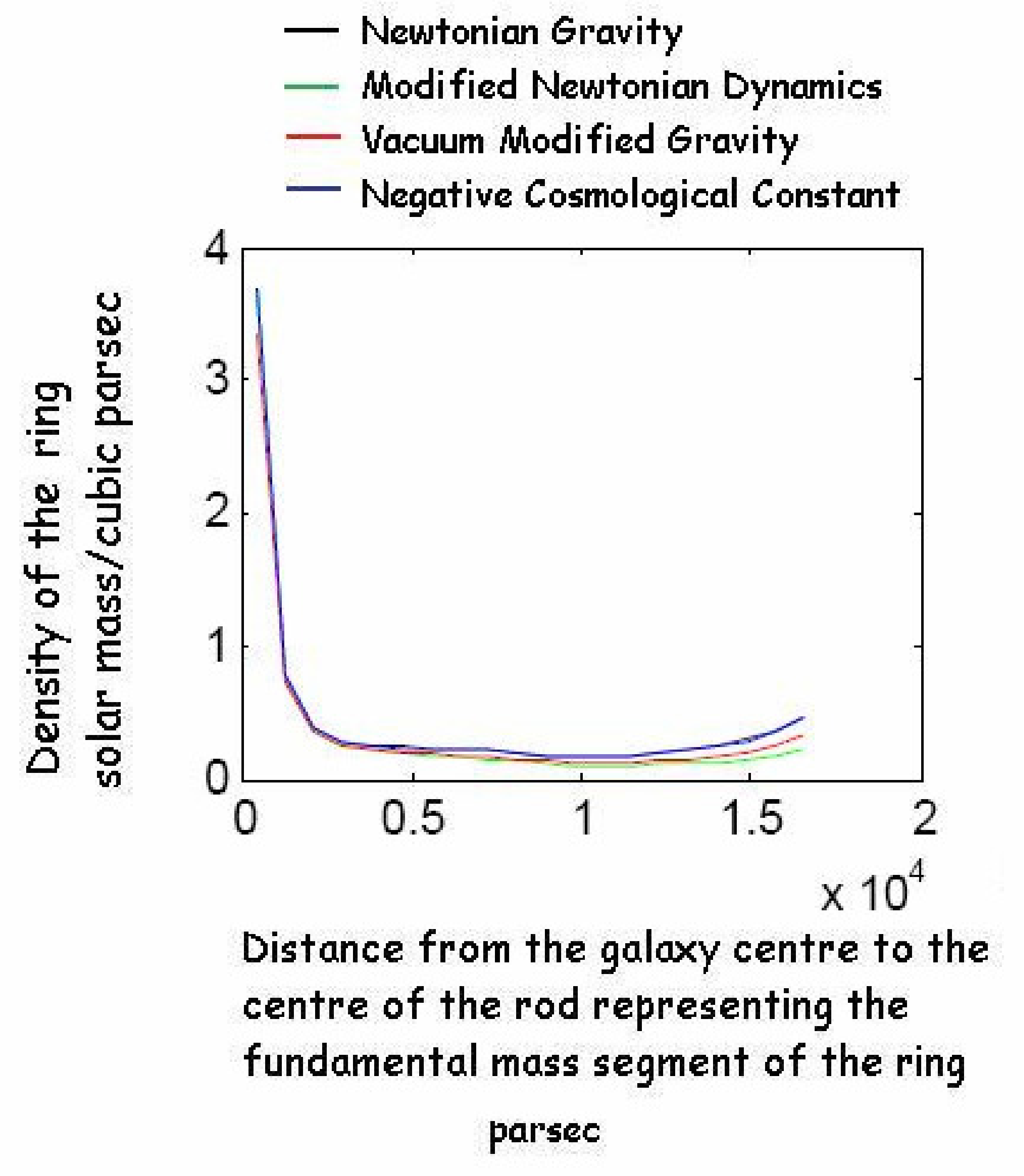

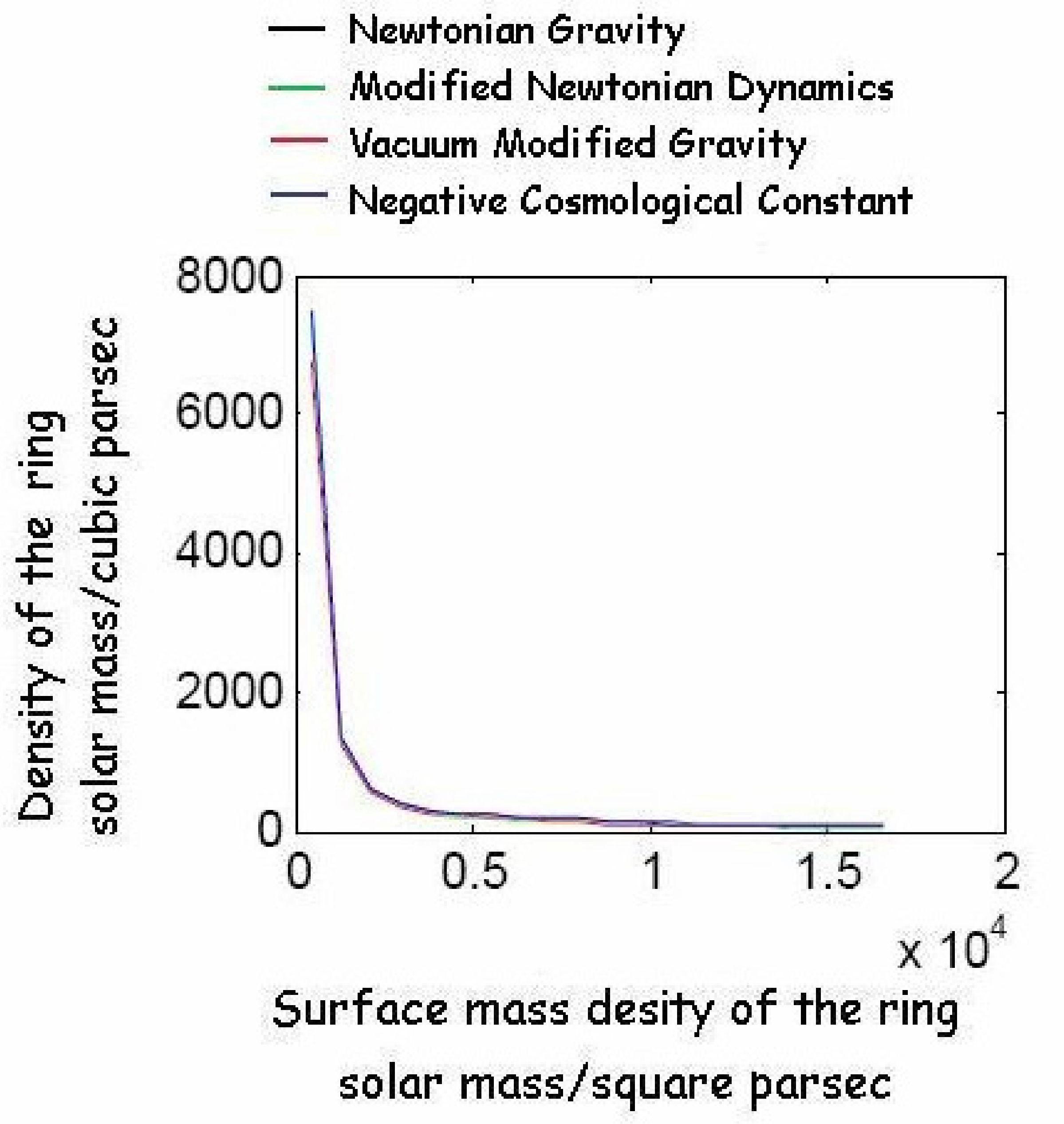
Newtonian Gravity
Modified Newtonian Dynamics
Vacuum Modified Gravity
Negative Cosmological Constant
8000
6000
4000
2000
0
Density of the ring
solar mass/cubic parsec
0
0.5
1
1.5
2
x 10^4
Surface mass desity of the ring
solar mass/square parsec

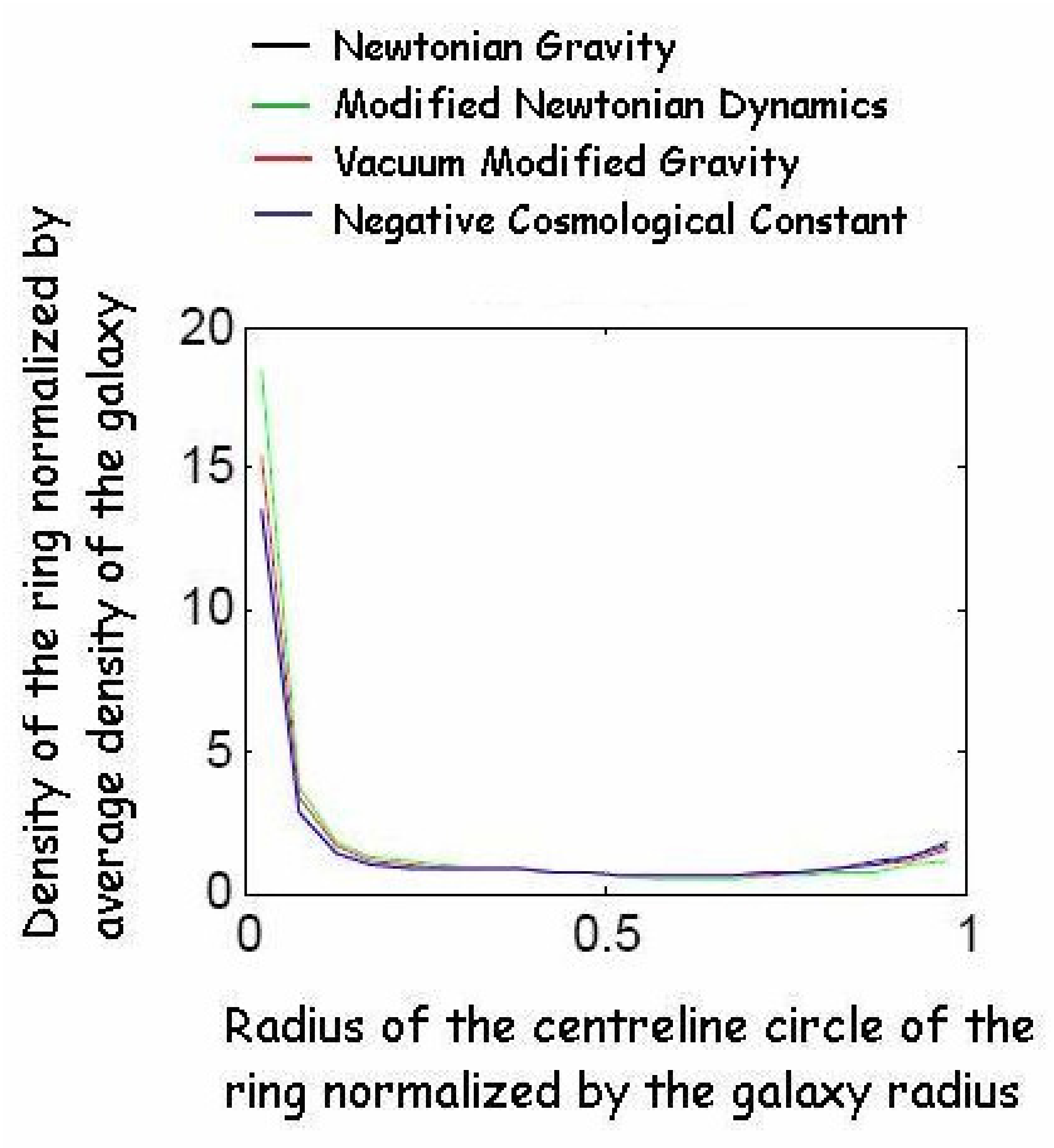
Newtonian Gravity
Modified Newtonian Dynamics
Vacuum Modified Gravity
Negative Cosmological Constant
Density of the ring normalized by average density of the galaxy
20
15
10
5
0
0
0.5
1
Radius of the centreline circle of the ring normalized by the galaxy radius

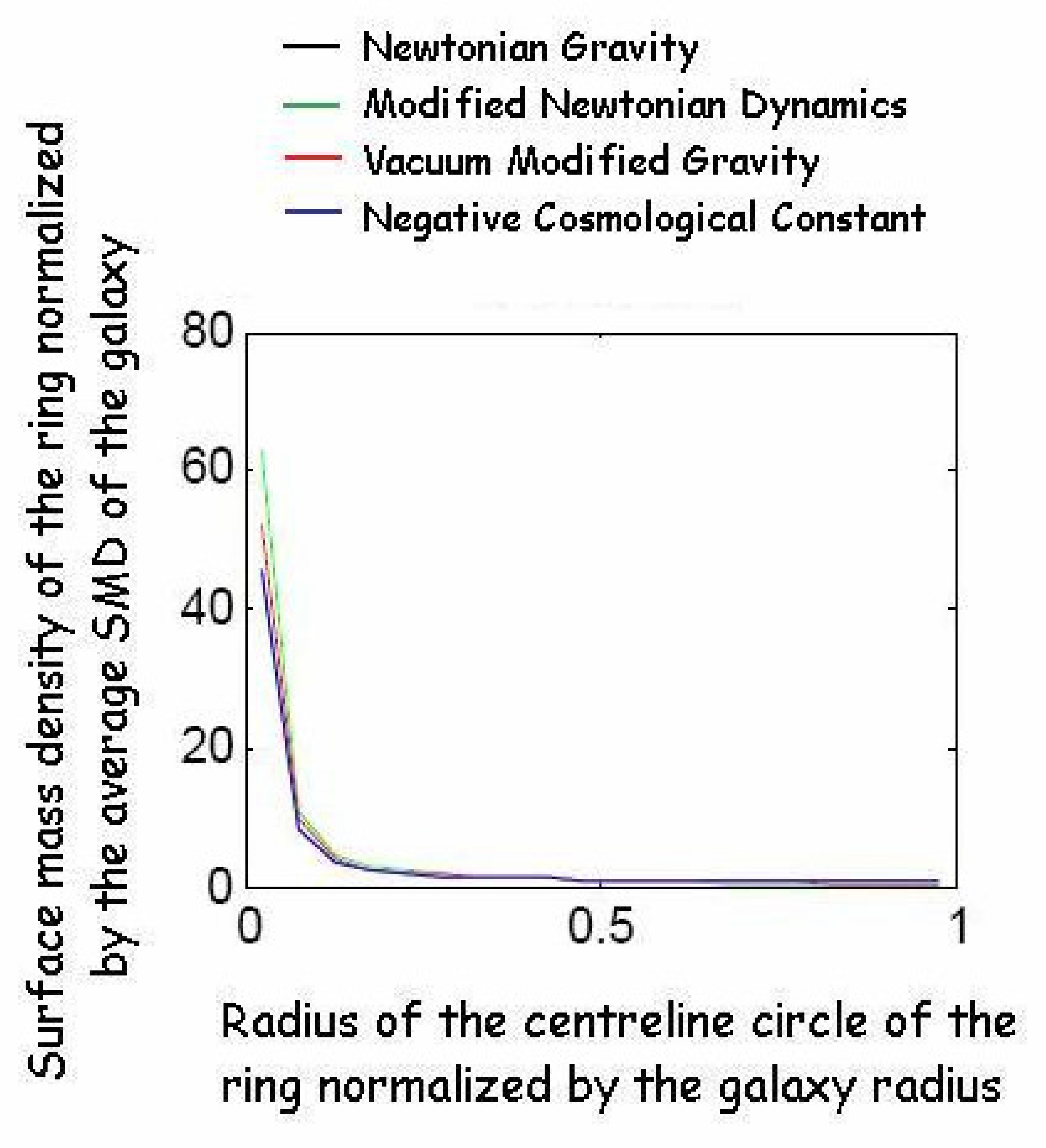
Newtonian Gravity
Modified Newtonian Dynamics
Vacuum Modified Gravity
Negative Cosmological Constant
80
60
40
20
0
0
0.5
1
Surface mass density of the ring normalized by the average SMD of the galaxy
Radius of the centreline circle of the ring normalized by the galaxy radius

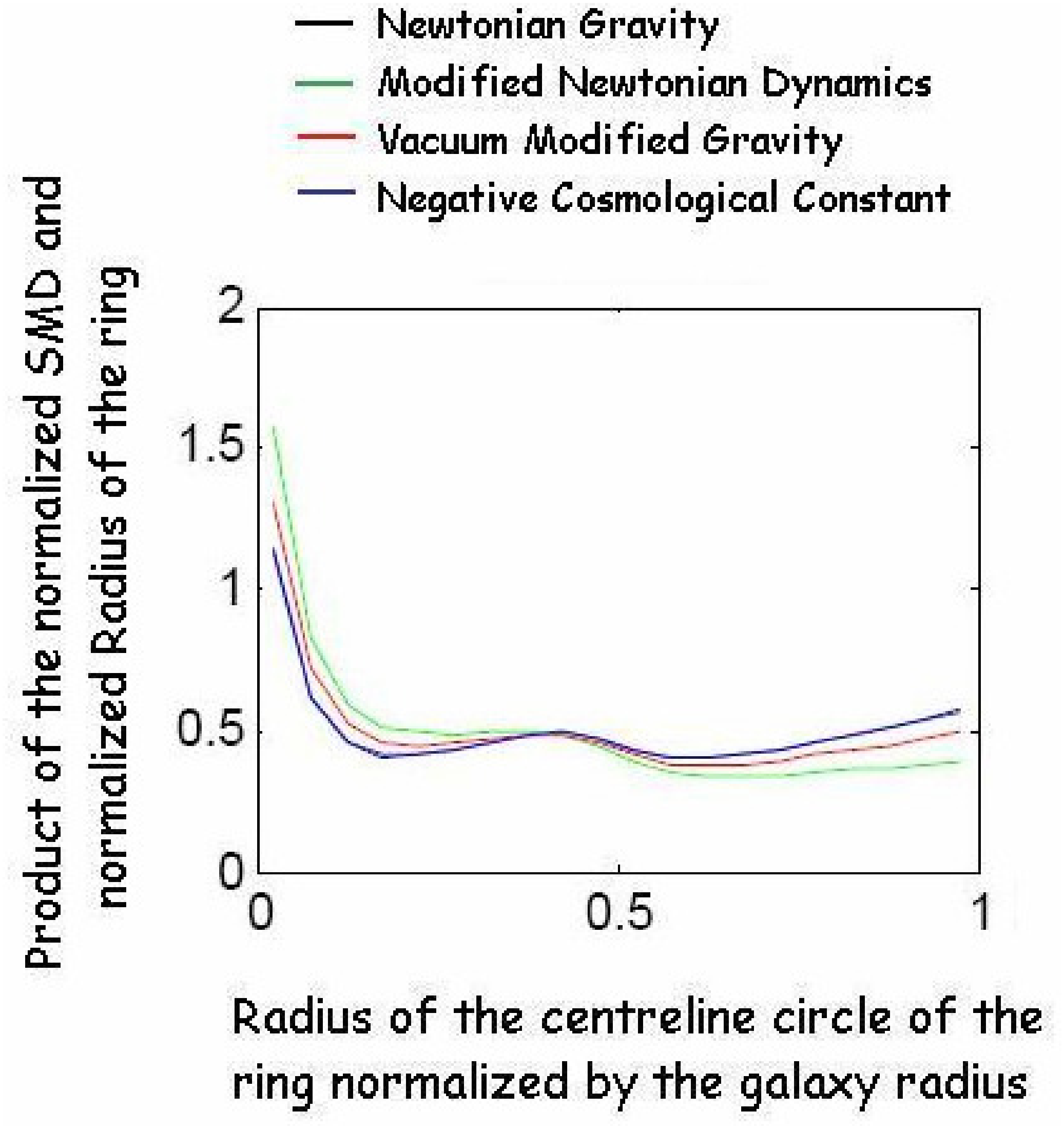
Newtonian Gravity
Modified Newtonian Dynamics
Vacuum Modified Gravity
Negative Cosmological Constant
Product of the normalized SMD and normalized Radius of the ring
2
1.5
1
0.5
0
0
0.5
1
Radius of the centreline circle of the ring normalized by the galaxy radius

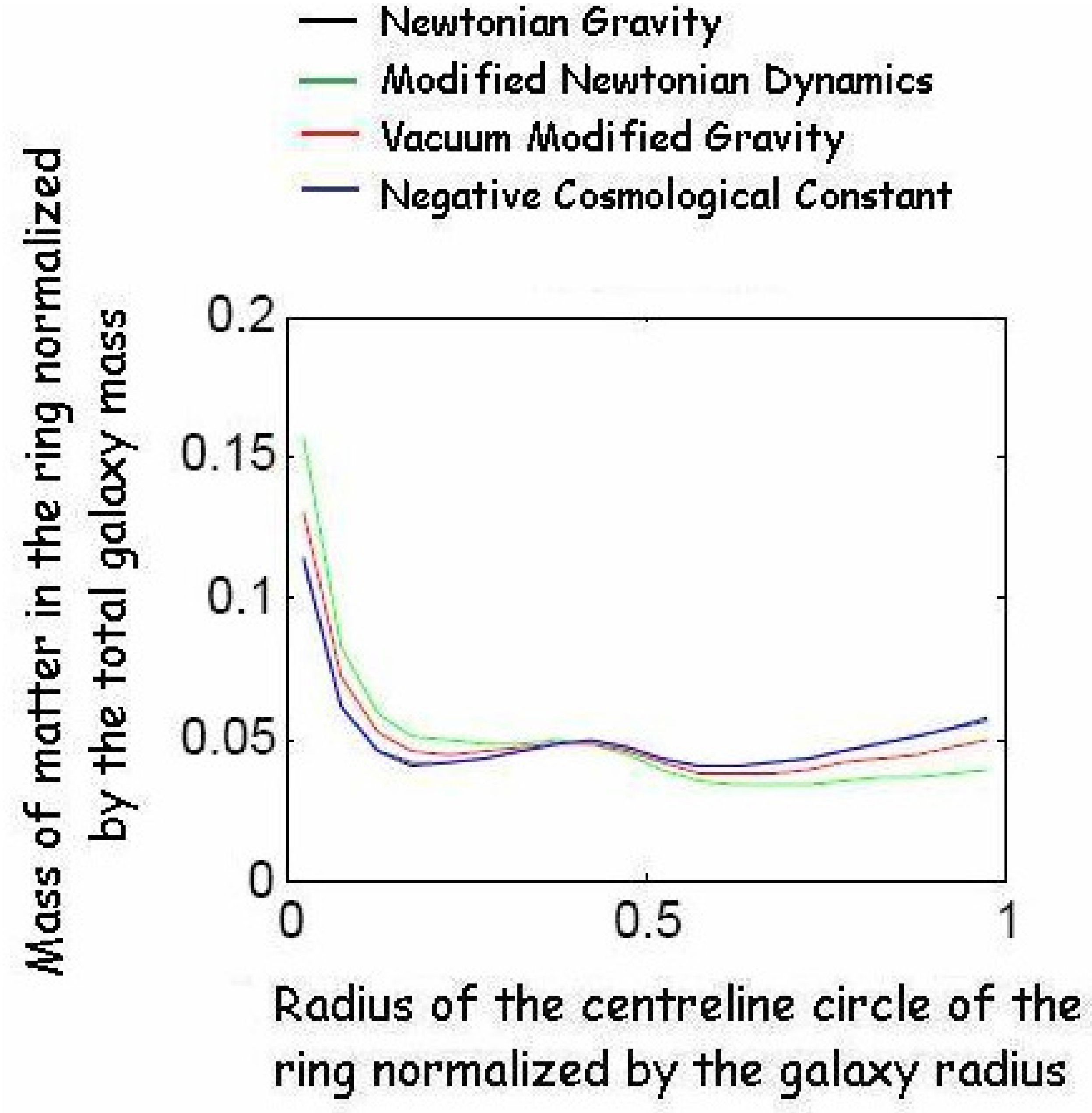

Newtonian Gravity
Modified Newtonian Dynamics
Vacuum Modified Gravity
Negative Cosmological Constant
Mass of matter in the ring normalized by the total galaxy mass
0.2
0.15
0.1
0.05
0
0
0.5
1
Radius of the centreline circle of the ring normalized by the galaxy radius

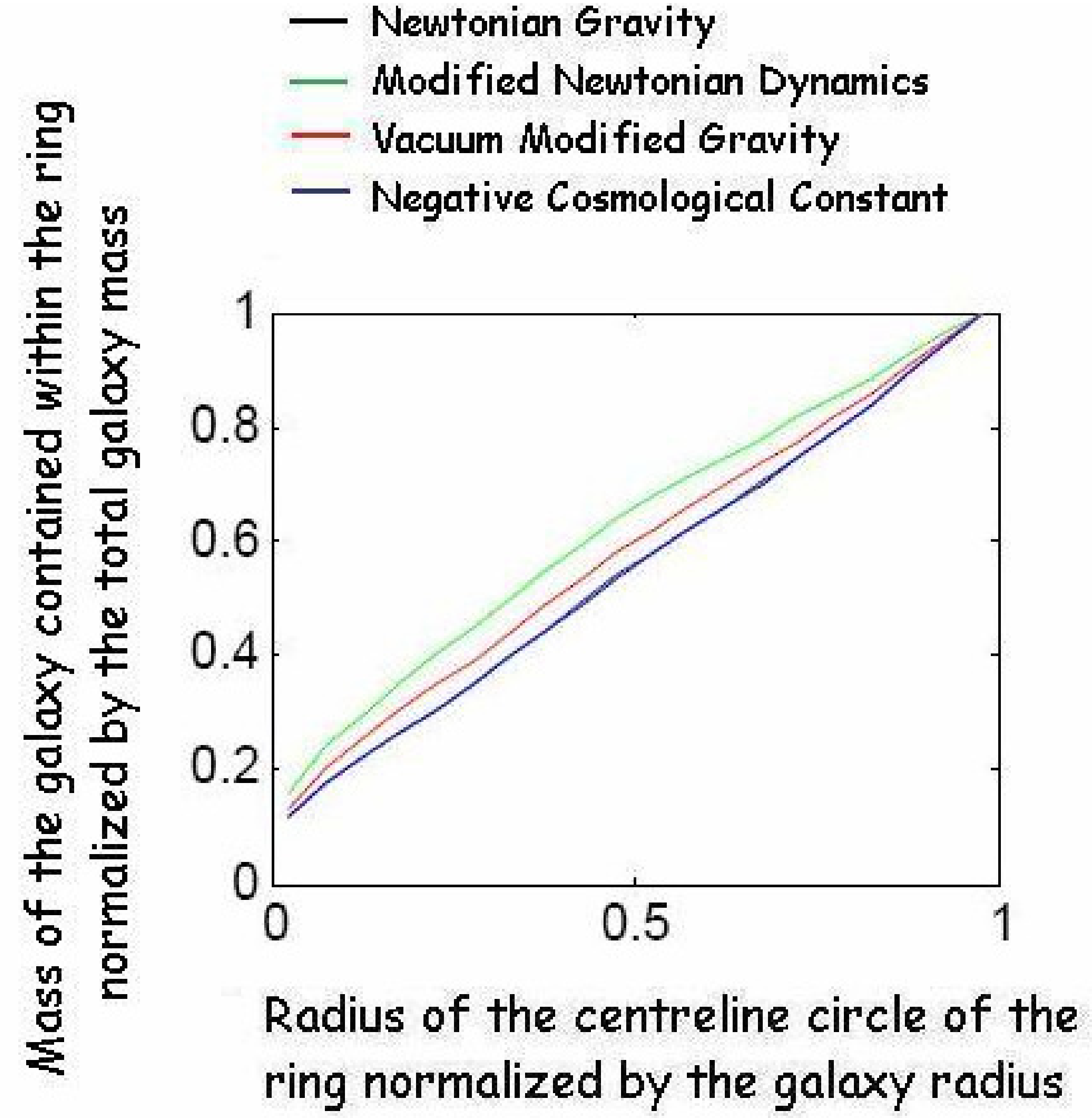
Newtonian Gravity
Modified Newtonian Dynamics
Vacuum Modified Gravity
Negative Cosmological Constant
Mass of the galaxy contained within the ring normalized by the total galaxy mass
Radius of the centreline circle of the ring normalized by the galaxy radius
0
0.2
0.4
0.6
0.8
1
0
0.5
1

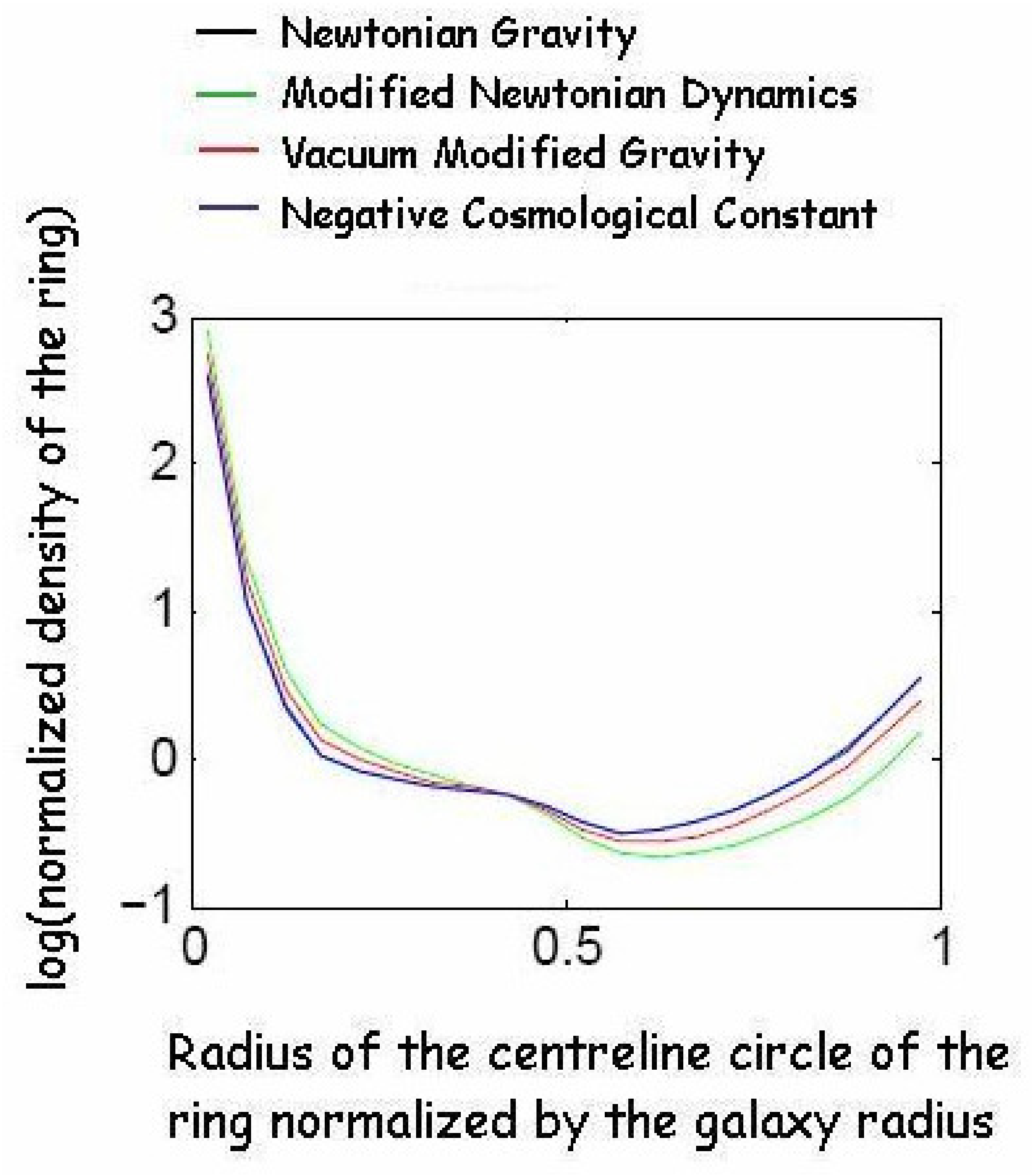
Newtonian Gravity
Modified Newtonian Dynamics
Vacuum Modified Gravity
Negative Cosmological Constant
log(normalized density of the ring)
3
2
1
0
−1
0
0.5
1
Radius of the centreline circle of the ring normalized by the galaxy radius

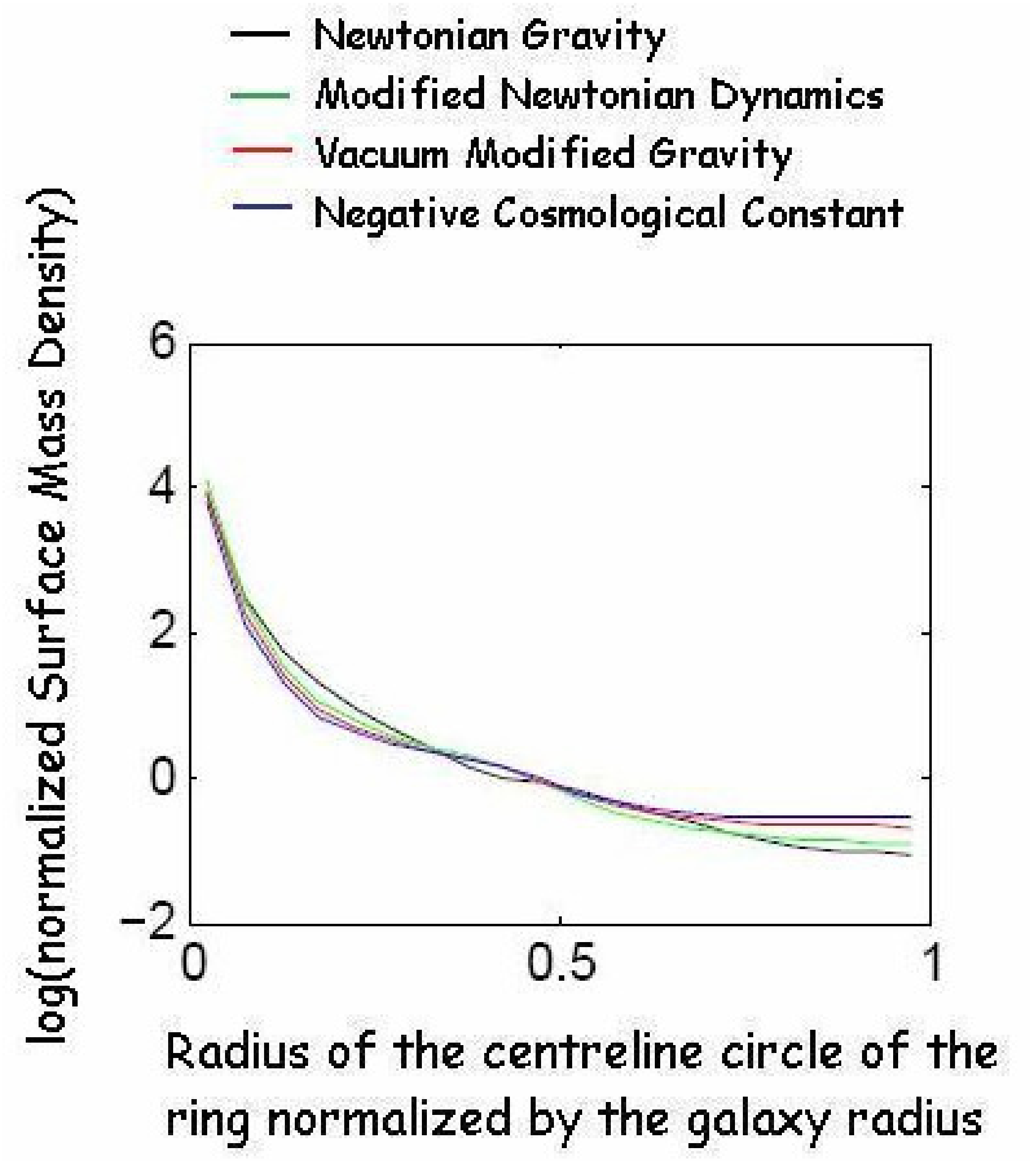
Newtonian Gravity
Modified Newtonian Dynamics
Vacuum Modified Gravity
Negative Cosmological Constant
log(normalized Surface Mass Density)
6
4
2
0
-2
0
0.5
1
Radius of the centreline circle of the ring normalized by the galaxy radius

## Milky Way (table)

| 20 Rings 17000 pc | Total Volume ($pc^3$) | Total Mass (solar mass, msuns) | Average Density (msuns/$pc^3$) | Average Surface Mass Density (msuns/$pc^2$) | Keplerian Velocity at Galaxy Rim (Kms/sec) |
|---|---|---|---|---|---|
| Newtonian Dyynamics | 5.4211 e+011 | 1.4955 E+011 | 0.2759 | 164.7222 | 194.5070 |
| Newtonian Dynamics with Negative Cosmological constant | 5.4211 e+011 | 1.4842 E+011 | 0.2738 | 163.4686 | 193.7654 |
| MoND | 5.4211 e+011 | 1.0694 E+011 | 0.1973 | 117.7889 | 164.4793 |
| Vacuum Modified Gravity | 5.4211 e+011 | 1.1782 E+011 | 0.2173 | 129.7735 | 172.6443 |

**NGC 3198 (30 kpc)**

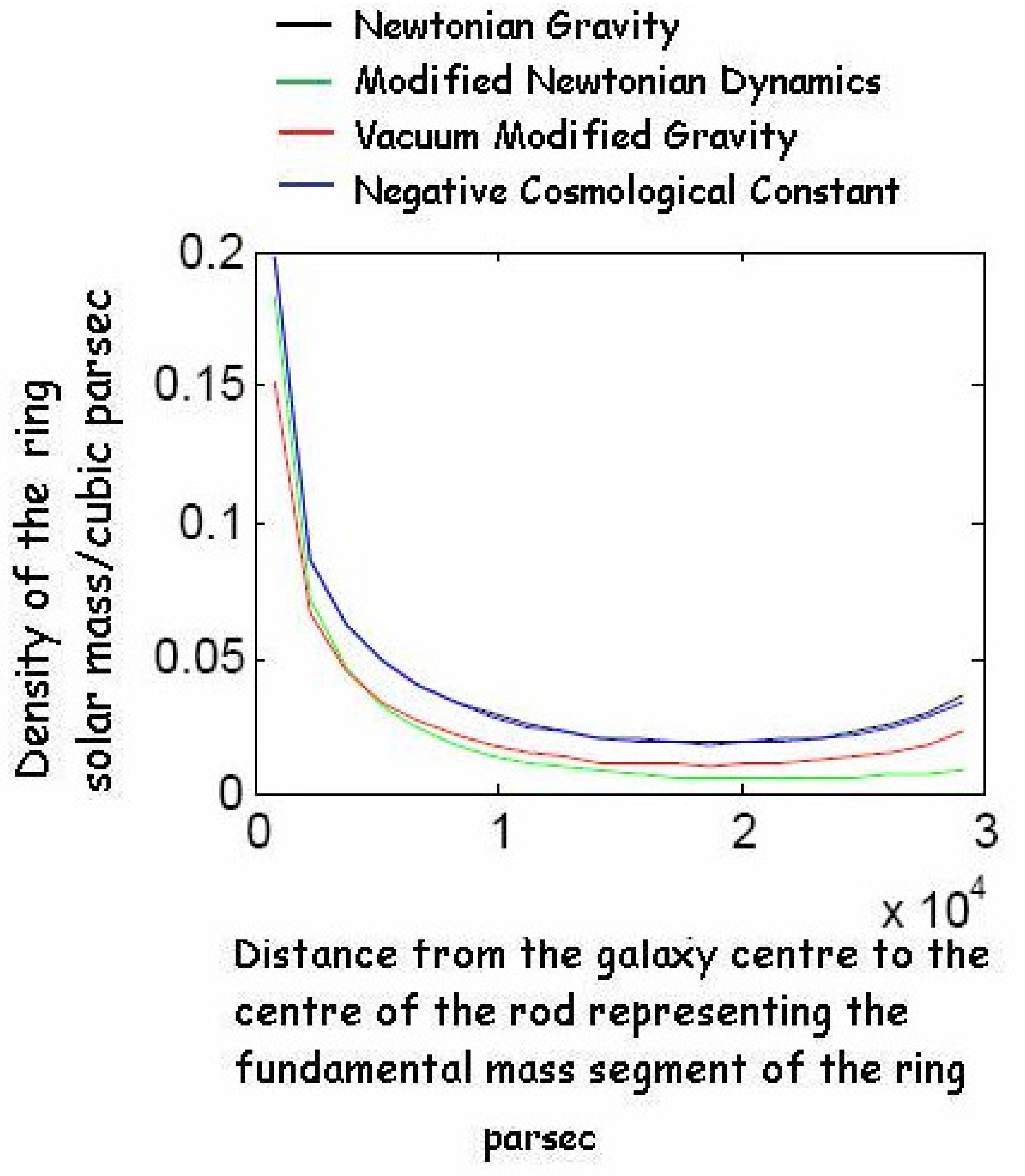

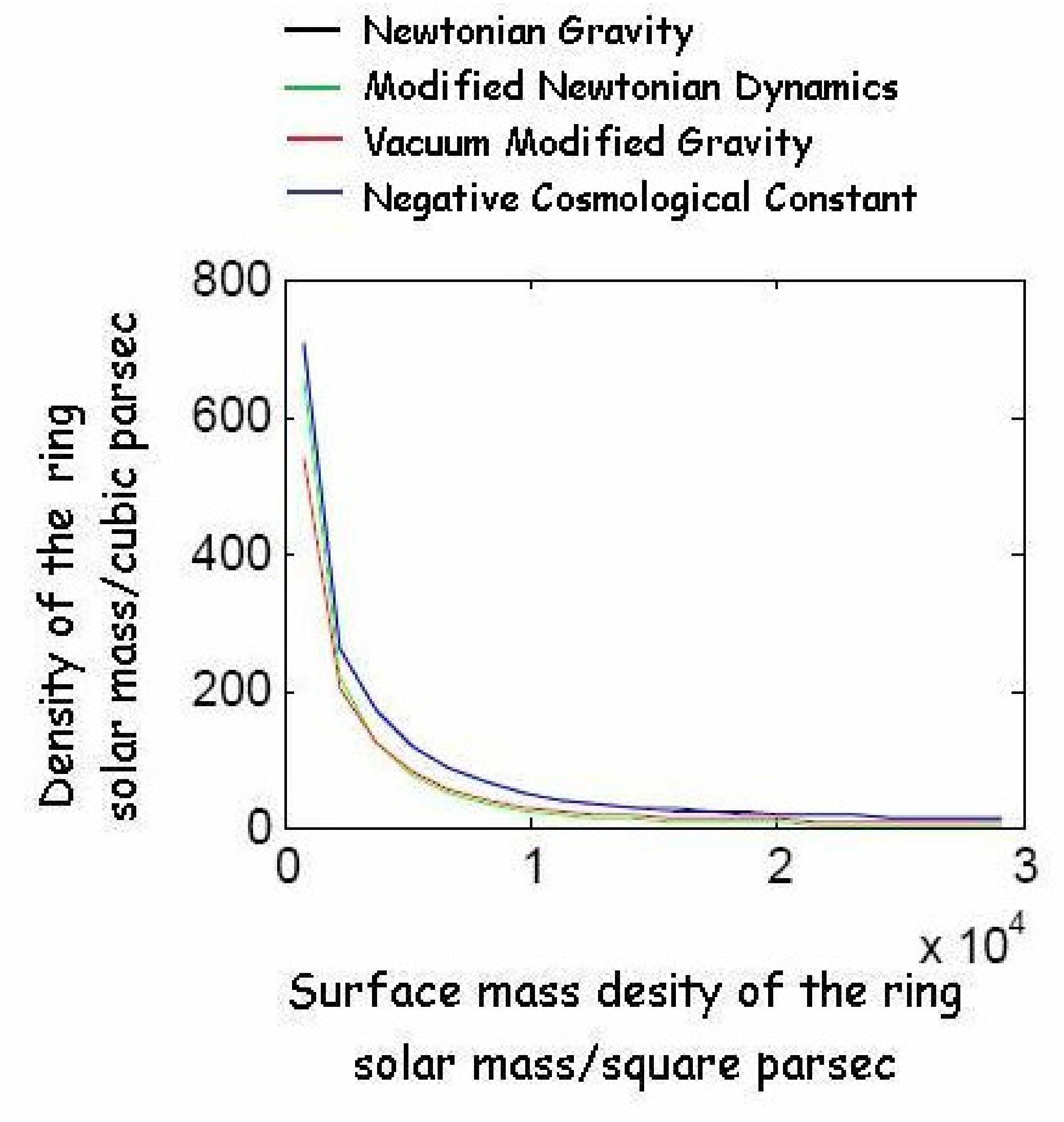
Newtonian Gravity
Modified Newtonian Dynamics
Vacuum Modified Gravity
Negative Cosmological Constant
800
600
400
200
0
0
1
2
3
x 10^4
Density of the ring
solar mass/cubic parsec
Surface mass desity of the ring
solar mass/square parsec

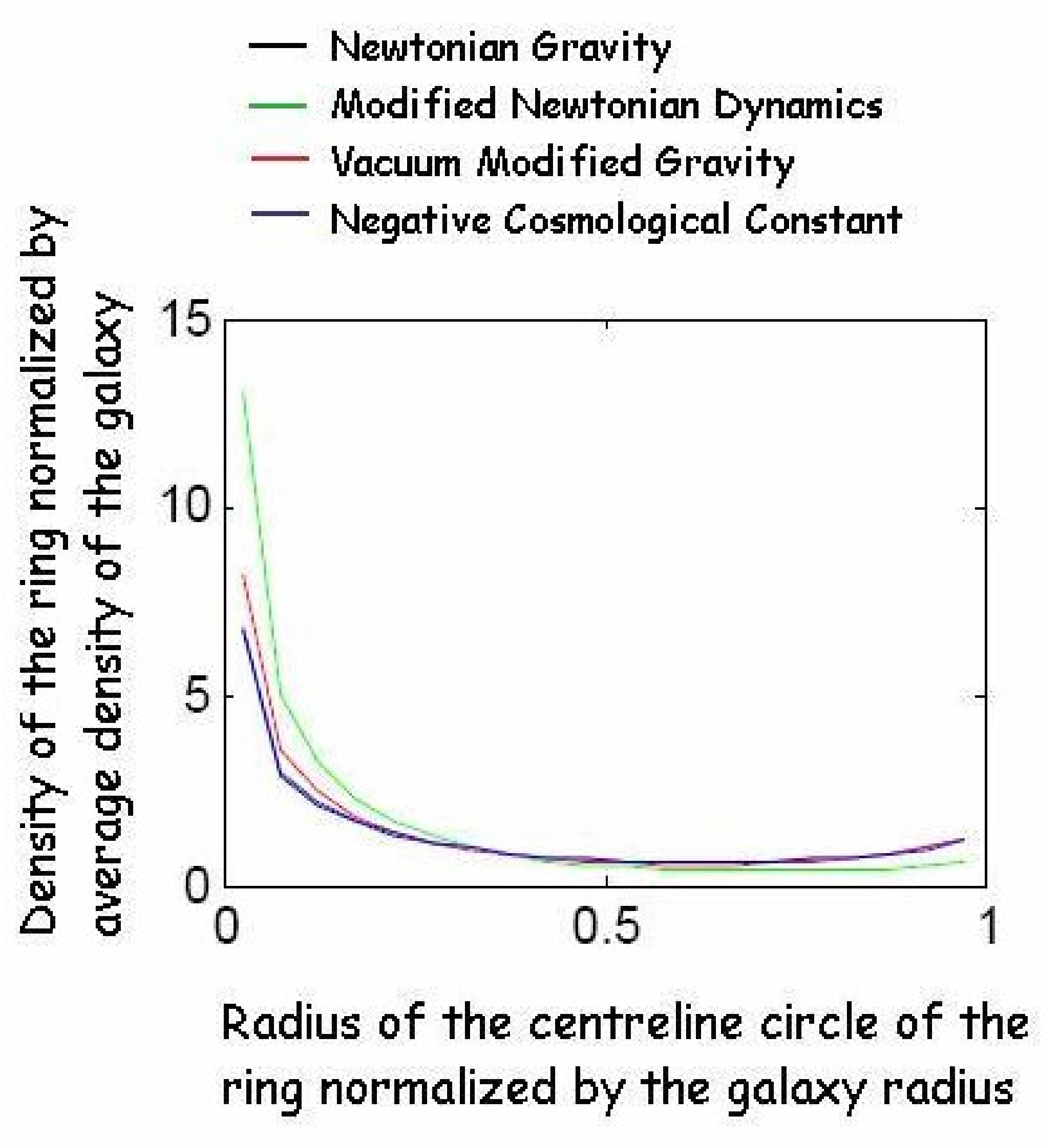
Newtonian Gravity
Modified Newtonian Dynamics
Vacuum Modified Gravity
Negative Cosmological Constant
Density of the ring normalized by average density of the galaxy
15
10
5
0
0
0.5
1
Radius of the centreline circle of the ring normalized by the galaxy radius

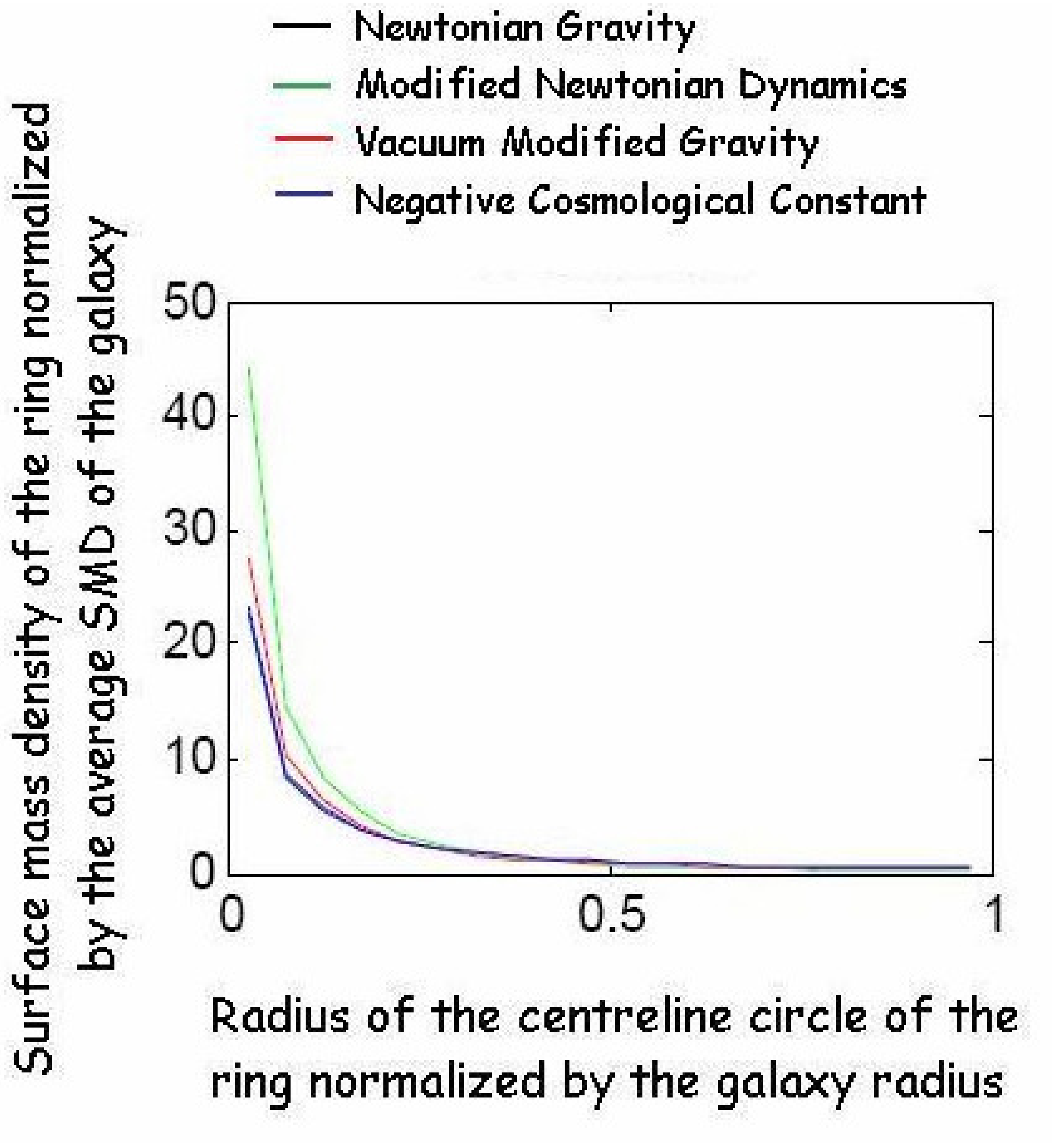

Newtonian Gravity
Modified Newtonian Dynamics
Vacuum Modified Gravity
Negative Cosmological Constant
Surface mass density of the ring normalized by the average SMD of the galaxy
50
40
30
20
10
0
0
0.5
1
Radius of the centreline circle of the ring normalized by the galaxy radius

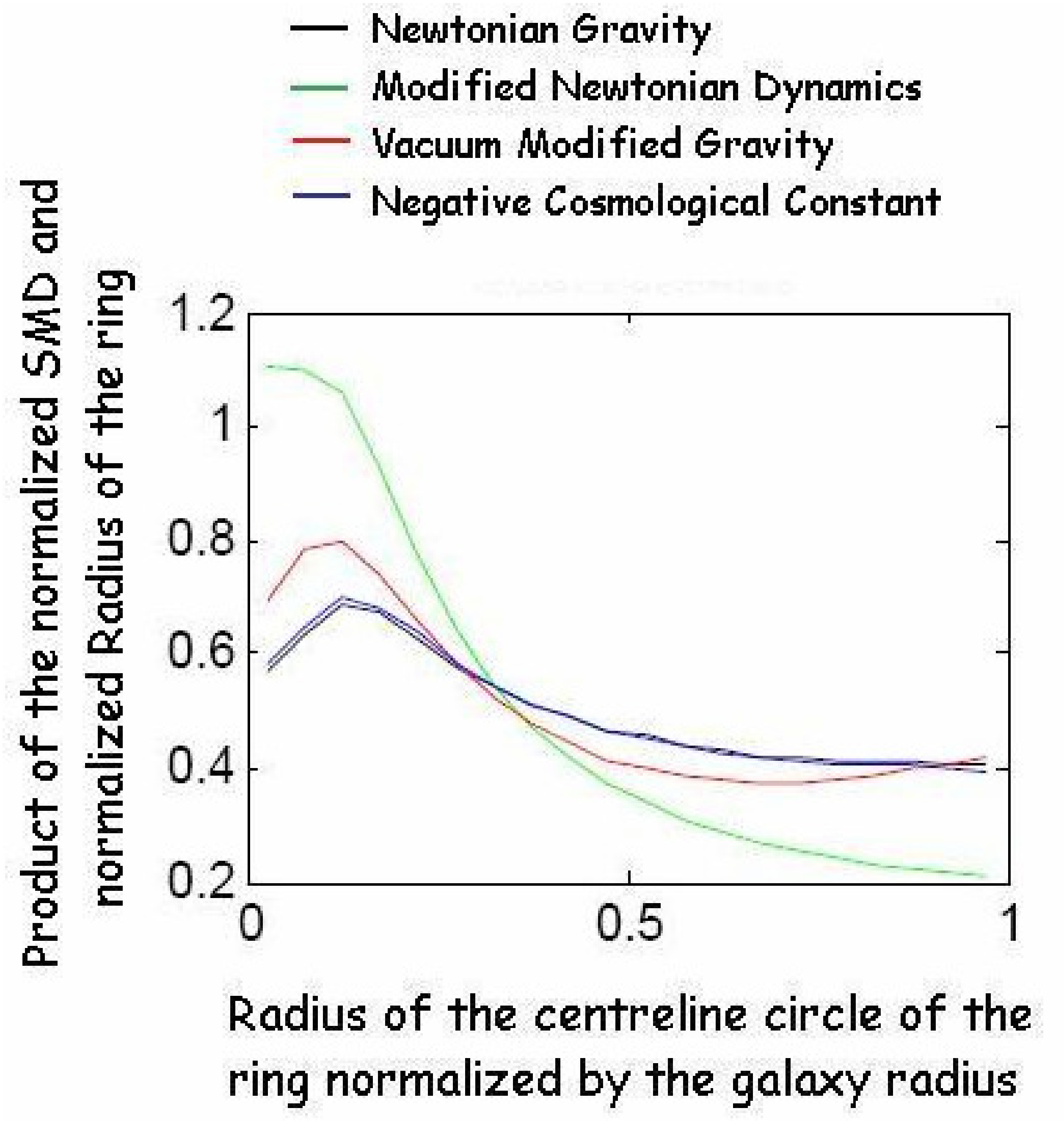

Newtonian Gravity
Modified Newtonian Dynamics
Vacuum Modified Gravity
Negative Cosmological Constant
Product of the normalized SMD and normalized Radius of the ring
1.2
1
0.8
0.6
0.4
0.2
0
0.5
1
Radius of the centreline circle of the ring normalized by the galaxy radius

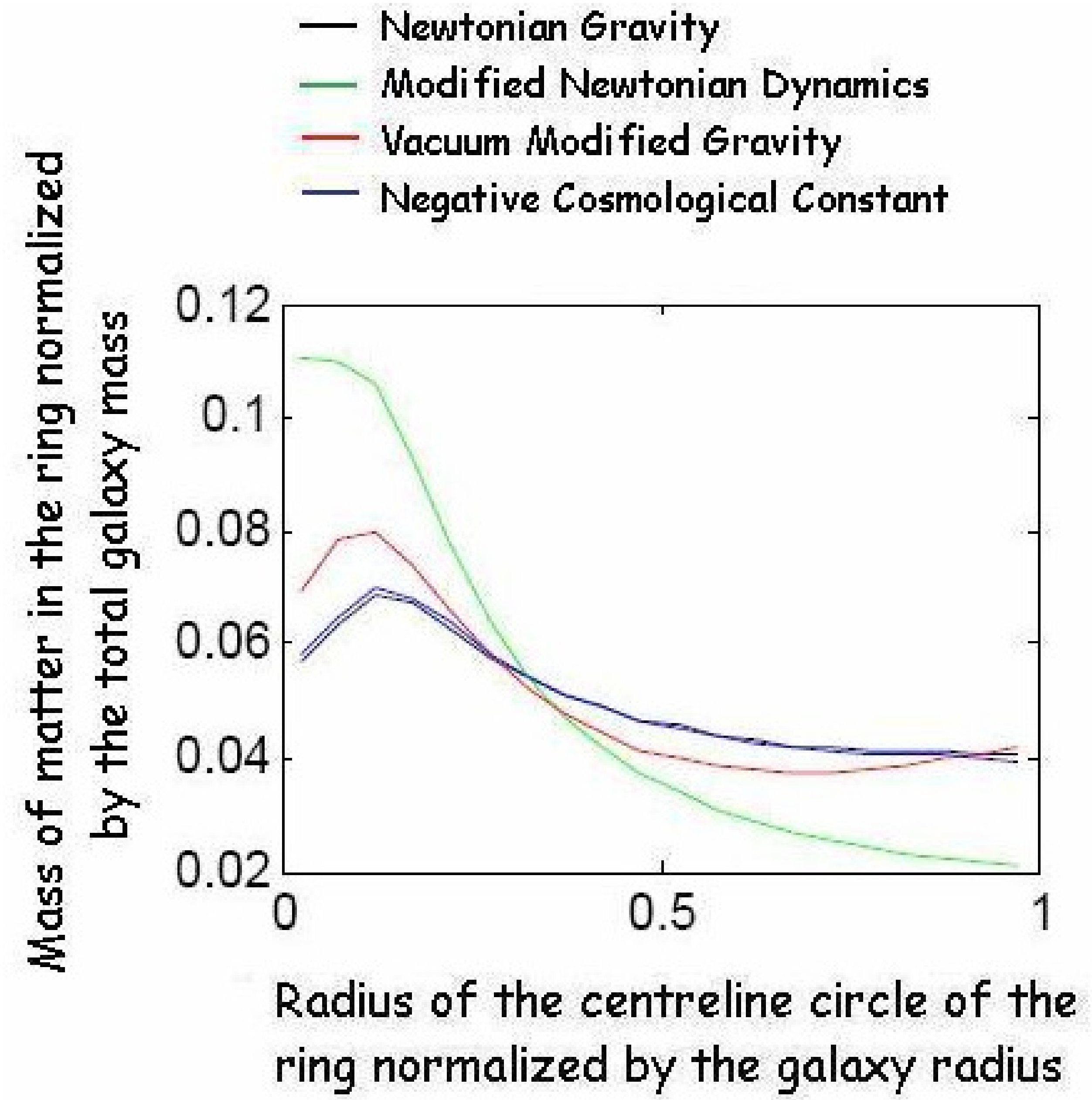
Newtonian Gravity
Modified Newtonian Dynamics
Vacuum Modified Gravity
Negative Cosmological Constant
Mass of matter in the ring normalized by the total galaxy mass
0.12
0.1
0.08
0.06
0.04
0.02
0
0.5
1
Radius of the centreline circle of the ring normalized by the galaxy radius

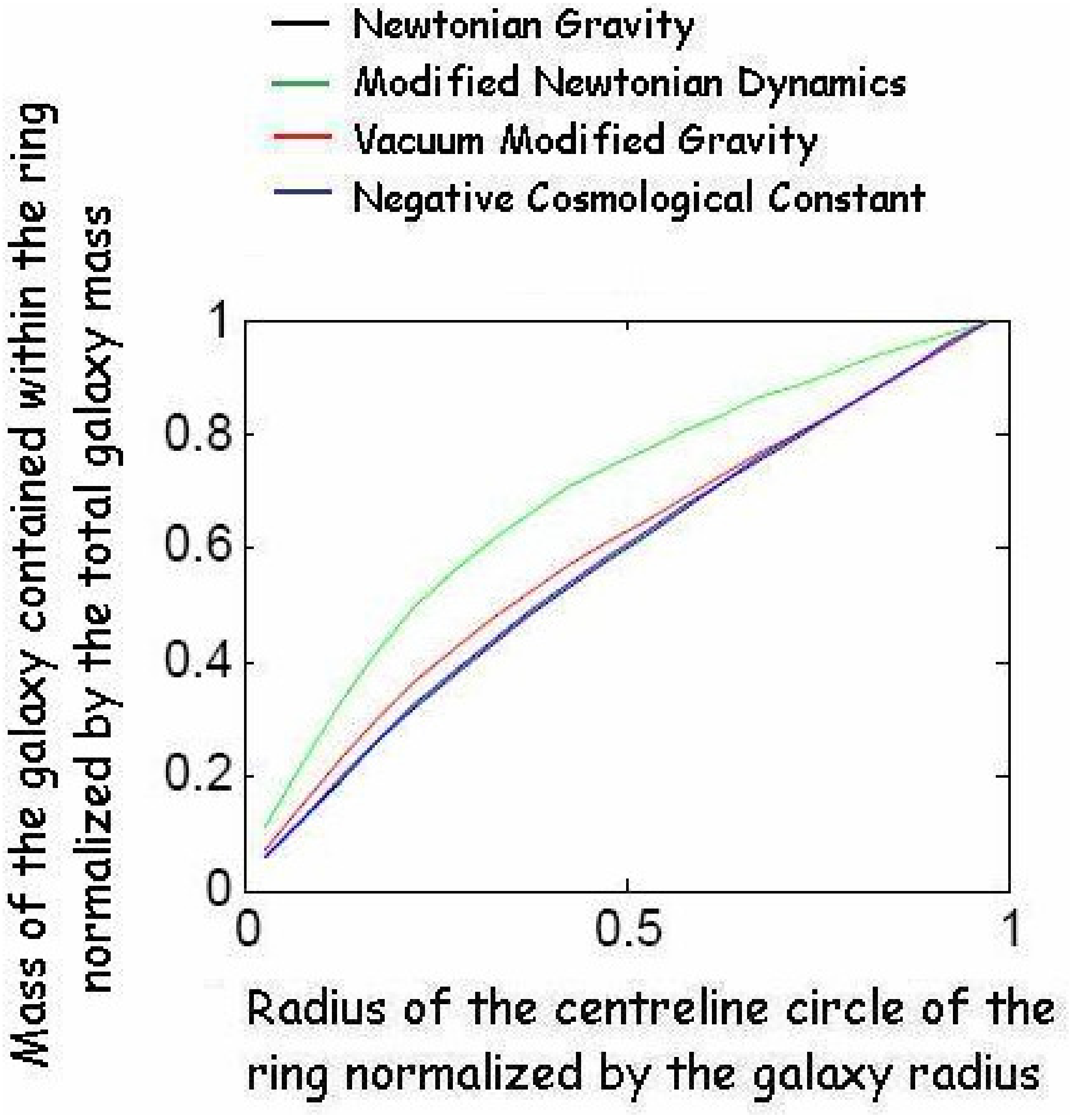
Newtonian Gravity
Modified Newtonian Dynamics
Vacuum Modified Gravity
Negative Cosmological Constant
Mass of the galaxy contained within the ring normalized by the total galaxy mass
0
0.2
0.4
0.6
0.8
1
0
0.5
1
Radius of the centreline circle of the ring normalized by the galaxy radius

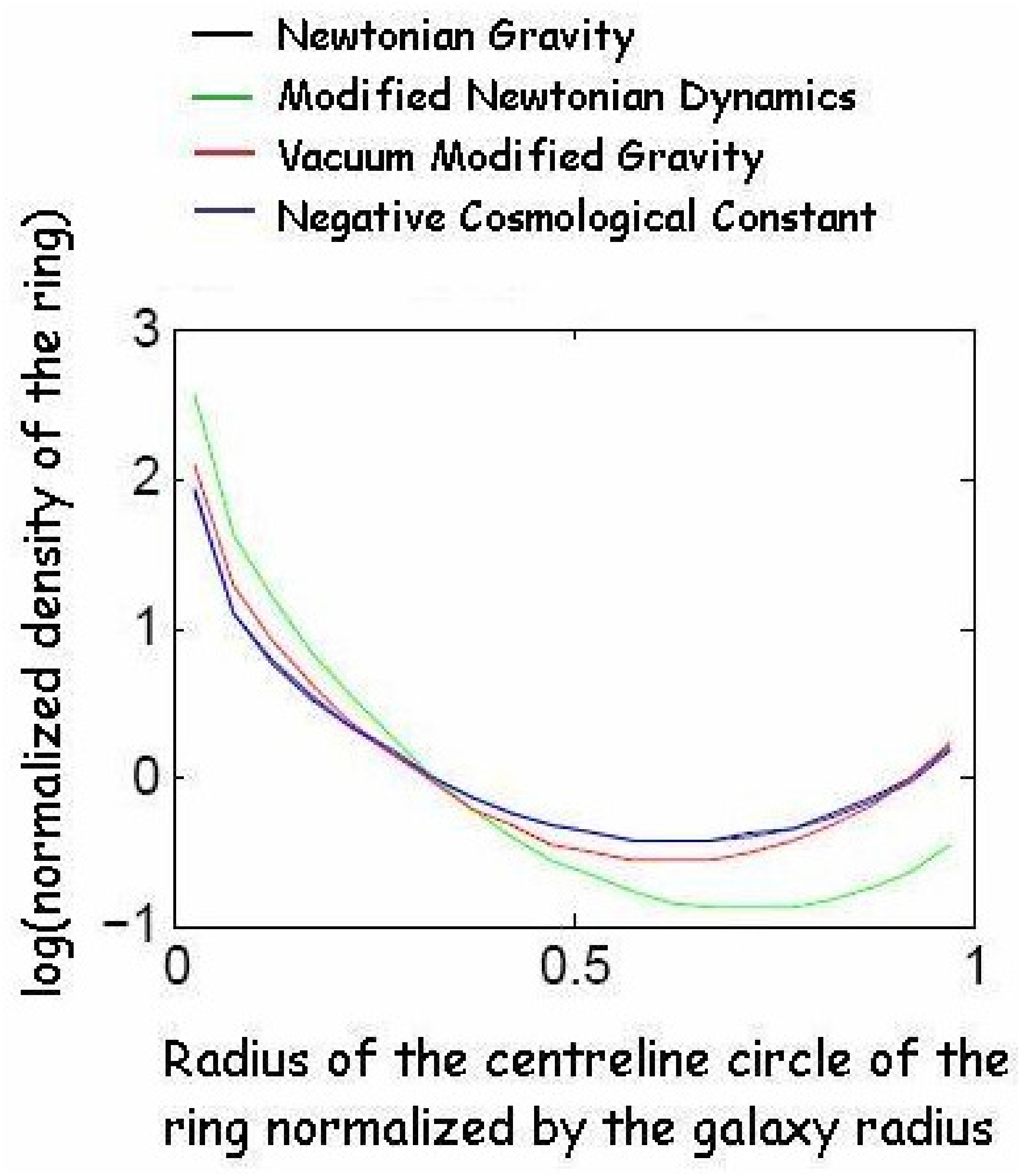
Newtonian Gravity
Modified Newtonian Dynamics
Vacuum Modified Gravity
Negative Cosmological Constant
log(normalized density of the ring)
3
2
1
0
−1
0
0.5
1
Radius of the centreline circle of the ring normalized by the galaxy radius

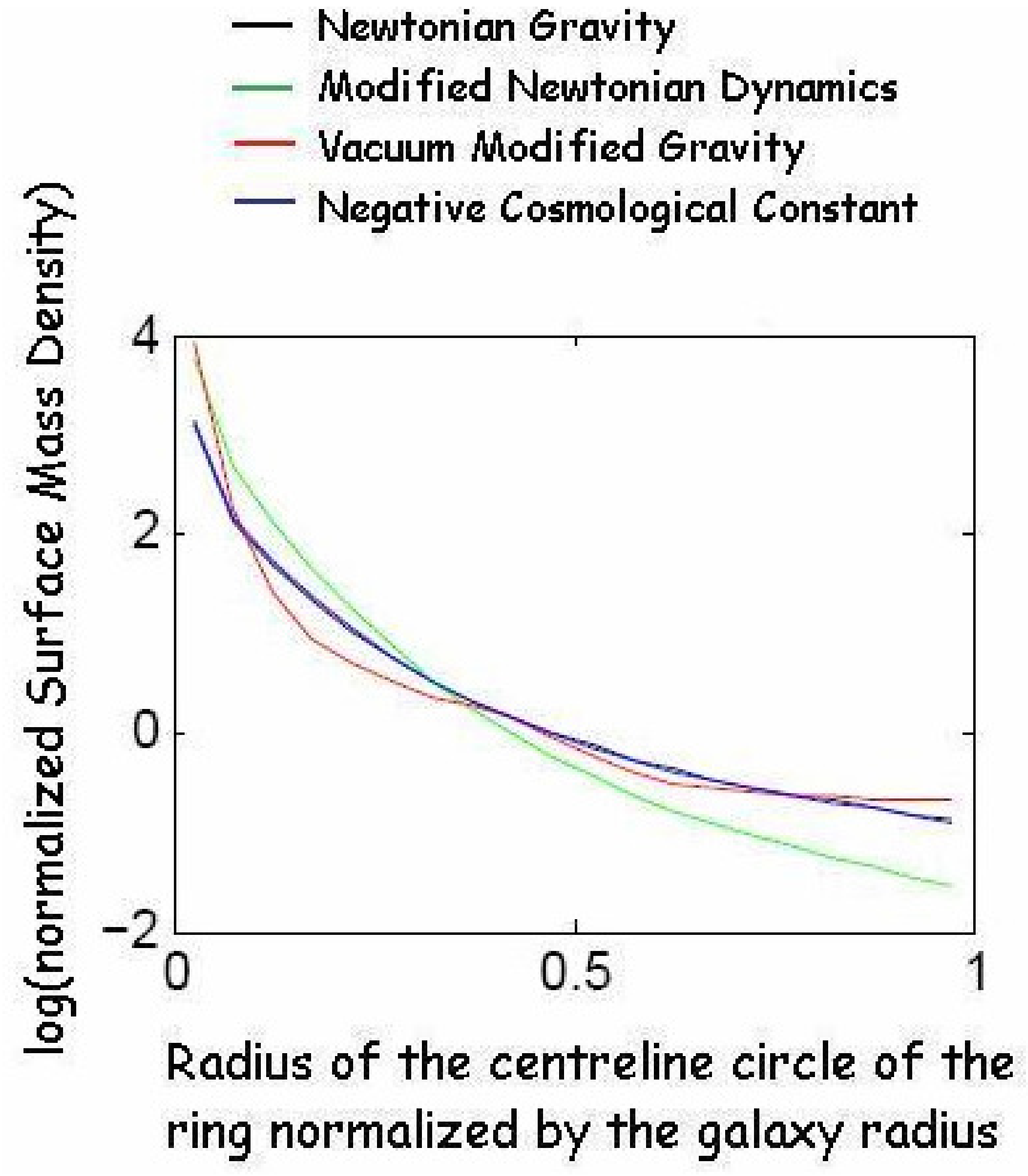
Newtonian Gravity
Modified Newtonian Dynamics
Vacuum Modified Gravity
Negative Cosmological Constant
log(normalized Surface Mass Density)
4
2
0
-2
0
0.5
1
Radius of the centreline circle of the ring normalized by the galaxy radius

## NGC 3198 (table)

| 20 Rings 30000 pc | Total Volume ($pc^3$) | Total Mass (solar mass, msuns) | Average Density (msuns/$pc^3$) | Average Surface Mass Density (msuns/$pc^2$) | Keplerian Velocity at Galaxy Rim (Kms/sec) |
|---|---|---|---|---|---|
| Newtonian Dyynamics | 2.9792 e+012 | 8.7450 E+010 | 0.0294 | 30.9290 | 111.9639 |
| Newtonian Dynamics with Negative Cosmological constant | 2.9792 e+012 | 8.5459 E+010 | 0.0287 | 30.2250 | 110.6824 |
| MoND | 2.9792 e+012 | 4.1890 E+010 | 0.0141 | 14.8154 | 77.4911 |
| Vacuum Modified Gravity | 2.9792 e+012 | 5.4961 e+010 | 0.0184 | 19.4384 | 88.7617 |

**Conclusions**

We can observe from the figures & tables, infer and conclude as follows:

1. As we move from smaller to larger galaxies, there is an increasing similarity between the mass and mass density profiles predicted as per different theories. For small galaxies, NGC6822 and LMC, there is a significant variation between what the different theories predict. As we move to Milky Way and NGC3198, we see much more similar looking curves and for UGC9133, the curves appear strikingly similar.
2. The well-studied luminosity profiles and mass-to-light ratios of galaxies suggest that there is an exponential decrease in the mass of visible matter as we move away from the centre of the galaxy. But, considering that ordinary matter in the outer reaches of a galaxy should be much cooler and may escape detection and the matter near the galactic centre should be very hot, we can probably explain why we see a very sharp fall in the mass and mass densities near the galactic centre and an almost constant profile as we move further outwards.
3. The mass and mass density profiles show an unexpected rising trend towards the outer edges of the galaxies and this trend is more prominent for smaller galaxies and the extent of the rise decreases as we move towards larger galaxies and becomes almost unnoticeable for very large galaxies.
4. For the larger galaxies, all the existing and proposed theories of gravity show nearly identical predictions, which appear intuitively satisfying considering what we already know regarding structures of galaxies. But for smaller galaxies, the predictions from different theories appear to vary significantly and also the results appear to contradict our expectations.
5. For example, MoND seems to give predictions very close to what is expected for galaxies of every size that we have examined whereas Newtonian theory gives bizarre predictions for small galaxies, though as we move to larger galaxies the predictions are closer to our expectations. Similar is the situation with Vacuum Modified Gravity.
6. Hence, what we may conclude is that there is a need to study smaller galaxies extensively in order to arrive at a conclusion regarding whether at all we need dark matter and whether MoND can be considered as a valid theory of gravity.
7. From the trend that is observed it appears that either MoND is correct or else there is a need to invoke dark matter along within Newtonian theory in order to explain the observations. And this dark matter, if there, seems to be concentrated near the outer edges of the galaxies with its gravitational effect on visible matter increasing as we move towards the galactic rim. And also the effect of dark matter compared to ordinary matter is more pronounced for smaller galaxies than the larger ones and is almost negligible for the very large ones.